\documentclass[prb,aps,preprint]{revtex4-1}
\usepackage{epsfig}
\usepackage{amsmath}
\draft
\begin{document}
\title{Path integral Monte Carlo simulations for rigid rotors and their application
to water}

\author{Eva G. Noya$^a$\footnote{eva.noya@iqfr.csic.es}, Luis M. Ses\'e$^b$, Rafael Ram\'{\i}rez$^c$, Carl McBride$^d$, Maria M. Conde$^d$ and Carlos Vega$^{d}$\footnote{cvega@quim.ucm.es}}
\affiliation{$^a$ Instituto de Qu\'{\i}mica F\'{\i}sica Rocasolano, Consejo Superior de Investigaciones Cient\'{\i}ficas, CSIC, Calle Serrano 119, 28006 Madrid, Spain }
\affiliation{$^b$Dept. Ciencias y T\'ecnicas Fisicoqu\'{\i}micas, Facultad de Ciencias, UNED, Paseo Senda del Rey 9, 28040 Madrid, Spain}
\affiliation{$^c$Instituto de Ciencia de Materiales, CSIC, Campus de Cantoblanco, 28049 Madrid, Spain}
\affiliation{$^d$ Departamento de Qu\'{\i}mica F\'{\i}sica,
Facultad de Ciencias Qu\'{\i}micas, Universidad Complutense de Madrid,
28040 Madrid, Spain}

\date{\today}

\begin{abstract}
In this work the path integral formulation for rigid rotors, proposed by M\"user and Berne [Phys. Rev. Lett. {\bf 77}, 2638 (1996)], is
described in detail. It is shown how this formulation can be used to perform Monte Carlo
simulations of water. Our numerical results show that whereas some properties of water 
can be  accurately reproduced using classical simulations with an empirical
potential which, implicitly, includes quantum effects, other properties can only
be  described quantitatively when quantum effects are explicitly incorporated.
In particular, quantum effects are extremely relevant when it comes to describing 
the equation of state of the ice phases at low temperatures, the structure
of the ices at low temperatures, and the heat capacity of both liquid water and the ice
phases. 
They also  play a minor role in the relative stability of the ice phases.
\end{abstract}
\maketitle

\section{Introduction}

	In 1948 Feynman proposed the path integral formulation of quantum
mechanics\cite{Feynman_PI,ebook_FeynmanHibbs}.
In the late seventies the works of Barker and of Chandler and Wolynes showed  
that this formulation
could be implemented in the statistical mechanical study of condensed matter by performing classical simulations of a modified
Hamiltonian\cite{JCP_1979_70_02914,JCP_1981_74_04078}.
It was demonstrated that the partition function of 
a quantum system of $N$ particles in its discretised form
is formally identical to that of a classical system consisting of $N$ ring polymers.
Thus a number of the techniques and methods that had already been derived for classical simulations could now
be adapted to perform quantum simulations. Since then path integral simulations
have been used to study the behaviour of a large number of systems. 
A detailed description of the path integral technique in statistical mechanics,  and its applications,
can be found in several reviews\cite{NATO_ASI_C_293_0155_photocopy,ARPC_1986_37_0401_nolotengo}.

	The path integral formulation can be implemented both in  Monte Carlo (MC)
and in molecular dynamics (MD) simulations. MC simulations only
provide thermodynamic properties. Although there has been some
controversy concerning the description of the correct dynamics, there
are now MD methods, such as centroid molecular 
dynamics\cite{JCP_1994_100_5106,JCP_1999_111_2371}
or the ring polymer molecular dynamics\cite{JCP_2004_121_3368} 
that approximate the quantum dynamics of the system.

	The path integral method has also been extended to study rigid
rotors. The first quantum MC simulations of rigid bodies were
performed  in the mid-eighties using a semi-classical 
approximation to derive the rotational contribution to the 
partition function\cite{CPL_1984_103_0357,JCP_1985_82_05164}. This semi-classical approximation was also
used within molecular dynamics in the beginning of the nineties\cite{JCP_1991_95_03728}.
A few years later
M\"user and Berne derived a propagator for the rotational
contribution to the density function for rigid bodies\cite{PRL_1996_77_002638,JPCM_1999_11_0R117}.
Centroid molecular dynamics has also been recently extended
to deal with rigid rotors\cite{JCP_2004_121_5992}.
These attempts to extend the path integral method to rigid bodies
are motivated by the desire to describe the quantum behaviour of molecules 
in a computationally efficient way, since various condensed matter properties 
are more likely  to be affected by inter-molecular vibrations
rather than  intra-molecular contributions\cite{JPCM_1999_11_0R117}. In this situation,
it seems reasonable to describe the molecules as rigid rotors and
ignore the intra-molecular vibrations. Usually intra-molecular vibrations 
exhibit high frequencies that require to use a large number of
`replicas' of the system, thus increasing considerably the computational
cost of quantum simulations. This problem can be alleviated by the
use of smart techniques such as the recently proposed ring polymer
contractor method\cite{JCP_2008_129_024105}. That said, it is our opinion
that the description of molecules as rigid bodies is particularly interesting in its self;
since it allows one to separate 
inter-molecular and intra-molecular quantum effects,  thus assigning their 
relative influence on the different properties of the condensed 
phases.

	In this work we shall describe the 
path integral formulation for rigid rotors proposed by M\"user and Berne.
Even though this formulation has already described in 
Refs. \onlinecite{PRL_1996_77_002638,JPCM_1999_11_0R117},
the implementation of a path integral Monte Carlo involves many technical
details which, in our opinion, are worth describing in detail. In addition,
we  show how this method can be applied to study quantum effects
for a range of properties of  water.

\section{Path integral for rigid rotors}

{\bf A. General path integral formulation}

The behaviour of a system of $N$ quantum non-spherical rigid 
particles can be described by 
the Schr\"odinger equation:
\begin{equation}
 \hat{H} \Psi_{\lambda} =  E_{\lambda} \Psi_{\lambda}
\end{equation}
where the Hamiltonian is given by:
\begin{equation}
\hat{H}=\hat{T}^{tra} + \hat{T}^{rot} + \hat{U}
\end{equation}
where $\hat{T}^{tra}$ is the translational kinetic energy operator
of the centre of mass of the molecules,
$\hat{T}^{rot}$ is the orientational kinetic energy operator
of the rigid molecules,
and $\hat{U}$ is the potential energy operator.
The partition function for this system at inverse temperature $\beta$
can be written as:
\begin{equation}
\label{eq_z1}
Z= \sum _{\lambda}\langle \Psi_{\lambda}|  e^{-\beta \hat{H}}|\Psi_{\lambda} \rangle = Tr  (e^{-\beta \hat{H}})=\sum _{\lambda} e^{-\beta E_{\lambda}}
\end{equation}
where $\Psi_{\lambda}$ and $E_{\lambda} $ are the eigenfunctions 
and the eigenvalues, respectively, of the Hamiltonian.
The solutions of the Schr\"odinger equation form a complete orthonormal basis 
(i.e. $\langle \Psi_{\lambda} | \Psi_{\lambda'} \rangle = \delta_{\lambda\lambda'}$).
The operator $\hat{\rho}=e^{-\beta\hat{H}}$ is defined by its
Taylor expansion:
\begin{equation}
\hat{\rho}=e^{-\beta\hat{H}}=\sum_{k=0}^{\infty} \frac{\hat{H}^k}{k!}
\end{equation}
The second term in Eq. \ref{eq_z1} implies first an integration over
the coordinates of the system $({\bf r},{\bf \omega})$ followed by a sum over
the states $\lambda$. Here ${\bf r}$ represents the set of Cartesian coordinates 
of the centre of mass of each particle and
${\bf \omega}$ are the set of Euler angles that define the orientation of each 
particle in the system.
Note that although we have used the eigenfunctions of the Hamiltonian 
to write the partition
function, any complete basis set can be used due to the fact that the trace is invariant
with respect to a change of basis\cite{macquarrie}.

We now introduce  the density function, which in the coordinate space 
is defined as:
\begin{equation}
\rho ({\bf r}, {\bf \omega} ,{\bf r'}, {\bf \omega'},\beta ) = 
\sum_{\lambda} \Psi_{\lambda}^* ({\bf r}, {\bf \omega} ) \exp (-\beta E_{\lambda} )
\Psi_{\lambda}({\bf r'},{\bf \omega'})
\end{equation}
The second term in Eq. \ref{eq_z1}  can be re-written  as:
\begin{equation}
Z= Tr ( e^{-\beta\hat{H}} ) = 
\int \sum_{\lambda} \Psi_{\lambda}^* ({\bf r}, {\bf \omega} ) \exp (-\beta E_{\lambda} )
\Psi_{\lambda}({\bf r},{\bf \omega}) d{\bf r}d{\bf \omega}=
\int \rho ({\bf r},{\bf \omega},{\bf r},{\bf \omega},\beta ) d{\bf r} d{\bf \omega}
\end{equation}
where  one first sums over the states $\lambda $ to obtain 
$\rho ({\bf r},{\bf \omega},{\bf r},{\bf \omega},\beta)$ and then
integrates over the coordinates.
One interesting property of the density function is that the product of two
density functions results in  another density function:
\begin{equation}
\int \rho({\bf r},{\bf \omega},{\bf r''},{\bf \omega''};\beta_1) 
\rho({\bf r''},{\bf \omega''},{\bf r'},{\bf \omega'};\beta_2) 
d{\bf r''} d{\bf \omega''}  = 
\rho({\bf r},{\bf \omega},{\bf r'},{\bf \omega'}; \beta_1+\beta_2)
\end{equation}
Naturally this can be generalised to any given number of terms.
Using this convolution 
property of the density function the partition function can be re-written as:
\begin{equation}
\label{eq_Z_1}
Z = \int 
\rho \left({\bf r}^{1},{\bf \omega}^1,{\bf r}^{2},{\bf \omega}^2;\frac{\beta}{P} \right) ...
\rho \left({\bf r}^{P-1},{\bf \omega}^{P-1},{\bf r}^P {\bf \omega}^P;\frac{\beta}{P} \right)
\rho \left({\bf r}^P,{\bf \omega}^P,{\bf r}^1,{\bf \omega}^1;\frac{\beta}{P} \right) 
d{\bf r}^{1} d{\bf \omega}^1 ... d{\bf r}^P d{\bf \omega}^P
\end{equation}

Usually it is not possible to solve the Schr\"odinger equation for a
system of $N$ rigid non-spherical interacting particles, so 
the eigenfunctions $\Psi_{\lambda}$ and eigenvalues $E_{\lambda}$ are unknown.
However, as mentioned before, the partition function (or what is the same,
the trace of the density matrix operator $\hat{\rho }$) is invariant independently
of the complete set of basis functions used. A convenient complete
basis set is the one formed by the eigenfunctions of position and
orientation operators which, for non-spherical particles, is given by:
\begin{equation}
|{\bf r}^t {\bf \omega}^t \rangle = \delta ({\bf r}-{\bf r}^t)
\delta ({\bf \omega}-{\bf \omega}^t)
\end{equation}
Therefore, we can write the partition function using the eigenfunctions
of the position and orientation operator by substituting the sum over
the states $\lambda $ by an integration over ${\bf r}$ and ${\bf \omega}$.
In this way, the partition function can be calculated by evaluating the density
function $\rho^{t,t+1}(\beta/P)$:
\begin{equation}
\rho^{t,t+1}(\beta/P) \equiv
\rho ({\bf r}^t,{\bf \omega}^t, {\bf r}^{t+1},{\bf \omega}^{t+1},\beta/P) = 
\left\langle {\bf r}^t {\bf \omega}^t \left\vert
\exp \left[ - \beta (\hat{T}^{tra} +\hat{T}^{rot} + \hat{U})/P\right]
\right\vert {\bf r}^{t+1} {\bf \omega}^{t+1} \right\rangle
\end{equation}
where $| {\bf r}^{t+1} {\bf \omega}^{t+1} \rangle$ are the eigenfunctions of the
position operator.
In principle, the exponential cannot be factorised because the potential
and kinetic energy do not commute. However, using the Trotter formula\cite{PAMS_1959_10_0545},
one can write the following exact formula in the limit of infinite $P$:
\begin{eqnarray}
& & \lim_{P\to\infty}  \rho ({\bf r}^t, {\bf \omega}^{t}, {\bf r}^{t+1}, {\bf \omega}^{t+1} ,\beta/P) =  \\
& & \left\langle {\bf r}^t {\bf \omega}^t \left\vert
\exp \left[ - \beta \hat{U}/2P \right]
\exp \left[ - \beta (\hat{T}^{tra} +\hat{T}^{rot})/P \right]
\exp \left[ - \beta \hat{U}/2P \right]
\right\vert {\bf r}^{t+1} {\bf \omega}^{t+1} \right\rangle \nonumber
\end{eqnarray}
where $P$ is  the Trotter number. Given the ring polymer isomorphism, $P$ is also known as
as the number of replicas or beads.

The operator $\hat{U}$ is diagonal in the coordinate representation, 
and for rigid rotors $\hat{T}^{tra}$ and 
$\hat{T}^{rot}$ commute. Therefore, for rigid rotors $\rho^{t,t+1} (\beta/P)$ can be
approximated as:
\begin{equation}
\rho^{t,t+1} (\beta/P) \approx \rho_{\mathrm{pot}}^{t,t+1} (\beta/P)
\rho_{\mathrm{tra}}^{t,t+1} (\beta/P)
\rho_{\mathrm{rot}}^{t,t+1} (\beta/P)
\label{eq_density_matrix}
\end{equation}
where 
\begin{equation}
\label{eq_rho_pot}
\rho_{\mathrm{pot}}^{t,t+1} (\beta/P)= \exp \left[ - \frac{\beta}{2P} 
             \left( U ({\bf r}^t,{\bf \omega}^t) +  U({\bf r}^{t+1},{\bf \omega}^{t+1}) \right) \right],
\end{equation}
\begin{equation}
\rho_{\mathrm{tra}}^{t,t+1} (\beta/P)= \left< {\bf r}^t \left\vert \exp \left( -\beta \hat{T}^{\mathrm{tra}} /P \right) \right\vert {\bf r}^{t+1} \right>,
\end{equation}
\begin{equation}
\rho_{\mathrm{rot}}^{t,t+1} (\beta/P)
=  \left<  {\bf\omega}^t \left\vert \exp \left( 
-\beta \hat{T}^{\mathrm{rot}} /P \right) \right\vert {\bf\omega}^{t+1} \right>.
\end{equation}
where $U({\bf r}^t,{\bf \omega}^t)$ is the potential energy of $N$ rotors whose positions and
orientations are specified by ${\bf r}^t,{\bf \omega}^t$.

It can be shown that the translational contribution is given by\cite{NATO_ASI_C_293_0155_photocopy}:
\begin{equation}
\rho_{\mathrm{tra}}^{t,t+1} (\beta/P)
= \left( \frac{MP}{2 \pi \hbar^2 \beta} \right)^{3/2}  \exp \left[ -  
\sum_{i=1}^N  \frac{MP}{2 \hbar^2 \beta} ( {\bf r}_i^{t} -  {\bf r}_i^{t+1})^2 \right],
\label{eq_rho_trans}
\end{equation}
where $M$ is the rotor mass and ${\bf r}_i^t$ are the coordinates of
the centre of mass of replica $t$ of rotor $i$.

{\bf B. Rotational propagator for a free rotor}

	In order to evaluate the partition function for a 
system of $N$ free rotors
it is necessary to evaluate the rotational propagator:
\begin{equation}
\rho_{\mathrm{rot}}^{t,t+1} (\beta/P)=
\left\langle {\bf \omega}^t \left\vert \exp \left( - \frac{\beta}{P} \hat{T}_{rot} \right) \right\vert  {\bf \omega}^{t+1}
\right\rangle
\end{equation}
 which can be written as the product:
\begin{equation}
\left\langle {\bf \omega}^t \left\vert \exp\left( - \frac{\beta}{P} \hat{T}^{rot} \right) \right\vert {\bf \omega}^{t+1} \right\rangle
= \prod_{i=1}^N 
\left\langle {\bf \omega}_i^t \left\vert \exp \left( - \frac{\beta}{P} \hat{T}_i^{rot} \right) \right\vert {\bf \omega}_i^{t+1} \right\rangle
\label{equ_18}
\end{equation}
where $|\omega_i^{t}\rangle$ are the 
eigenfunctions of the orientation operator, which, as mentioned before, are
Dirac delta functions:
\begin{equation} 
|{\bf \omega}_i^{t}\rangle = \delta (\Omega-\Omega_i^t)
\end{equation}
Here $\Omega$ represents the three Euler angles 
$(\theta,\phi,\chi)$.
To simplify  the notation, we shall drop the subindex $i$, and from here
on focus our attention on a single free rotor.

The eigenfunctions of the angular position $|{\bf \omega}^t\rangle $ can be expanded
in a basis set of the eigenfunctions of the asymmetric top 
$| J M \hat{K} \rangle $ (the derivation of the eigenfunctions of the
asymmetric top are given in Appendix A):
\begin{equation}
|{\bf \omega}^{t}\rangle = \sum_{J} \sum_{M} \sum_{\hat{K}} 
\langle J M \hat{K} | {\bf \omega}^{t} \rangle | J M \hat{K} \rangle
\end{equation}
Using this expansion, the rotational propagator can  be rewritten as:
\begin{equation}
\left\langle {\bf \omega}^{t} \left\vert \exp \left( - \frac{\beta}{P} \hat{T}^{rot} \right) 
\sum_J \sum_M \sum_{\hat{K}} \left\langle J M \hat{K} \left. \right\vert {\bf \omega}^{t+1} \right\rangle 
\right\vert J M \hat{K} \right\rangle
\end{equation}
	As the functions $| J M \hat{K} \rangle $ are the eigenfunctions
of the Schr\"odinger equation for the asymmetric top, it follows that:

\begin{equation}
\left. \left. \exp \left( -\frac{\beta}{P} \hat{T}^{rot} \right) \right\vert J M \hat{K} \right\rangle =
\left. \left. \exp \left( -\frac{\beta}{P} E_{\hat{K}}^{(JM)} \right) \right\vert J M \hat{K} \right\rangle
\end{equation}
Using this, the rotational propagator can be written as:
\begin{equation}
\left\langle {\bf \omega}^t \left\vert \left( \sum_J \sum_M \sum_{\hat{K}} 
\exp \left( -\frac{\beta}{P} E_{\hat{K}}^{(JM)} \right)
\left\langle J M \hat{K} \left. \right\vert {\bf \omega}^{t+1} \right\rangle \right) \right\vert J M \hat{K} \right\rangle
\end{equation}
Reordering this expression one has:
\begin{equation}
\left\langle {\bf \omega}^{t} \left\vert \exp \left( -\frac{\beta}{P} \hat{T}^{rot} \right) \right\vert {\bf \omega}^{t+1}
\right\rangle = 
\sum_J \sum_M \sum_{\hat{K}} \left\langle {\bf \omega}^t \left. \right\vert J M \hat{K} \right\rangle
\exp \left( - \frac{\beta}{P} E_{\hat{K}}^{(JM)} \right) \left\langle J M \hat{K} \left. \right\vert
{\bf \omega}^{t+1} \right\rangle
\label{propagator}
\end{equation}

	The location of the laboratory frame defining the Euler angles
is arbitrary. For convenience we choose a laboratory frame such that the Euler angles
of replica $t$ are all zero (i.e., $\Omega^t = (0,0,0)$). 
This change leads to:
\begin{equation}
|{\bf \omega}^t \rangle = \delta (\Omega - \Omega^t ) = \delta (\Omega )
\end{equation}
and
\begin{equation}
|{\bf \omega}^{t+1} \rangle = 
\delta ( \tilde{\Omega}-\tilde{\Omega}^{t+1})
\end{equation}
A tilde is added to $\Omega$ (i.e., $\tilde{\Omega}$) in order to remind ourselves
that the Euler angles are defined in a laboratory frame in which the Euler angles
of bead $t$ are zero. Thus $\tilde{\Omega}^{t+1}$ are the
Euler angles of bead $t+1$ in this arbitrary frame.

	To simplify this expression further, the eigenfunctions of the
asymmetric top $|J M \hat{K} \rangle$ are expanded in a basis set
formed by the eigenfunctions of the symmetric top ($|J M K \rangle $):
\begin{equation}
| J M \hat{K} \rangle = \sum_{K} A_{\hat{K}K}^{JM} | J M K \rangle
\end{equation}

Eq. \ref{propagator} can now be re-written as:
\begin{eqnarray}
\left\langle {\bf \omega}^{t} \left\vert \exp \left( -\frac{\beta}{P} \hat{T}^{rot} \right) \right\vert {\bf \omega}^{t+1}
\right\rangle & = & 
\sum_J \sum_M \sum_{\hat{K}} \left[
\int \delta(\Omega) \sum_K A_{\hat{K}K}^{(JM)} 
| J M K \rangle 
d\Omega
\right] \exp \left( - \frac{\beta}{P} E_{\hat{K}}^{(JM)} \right) \nonumber \\
& & \left[ \int \left(\sum_K A_{\hat{K}K}^{(JM)} | J M K \rangle \right)^*
\delta (\tilde{\Omega}-\tilde{\Omega}^{t+1}) d\tilde{\Omega} \right] \nonumber \\
& = & \sum_J \sum_M \sum_{\hat{K}} \left( \sum_K A_{\hat{K}K}^{(JM)} 
\Psi_{JMK} (0)
\right)  \exp \left( -\frac{\beta}{P} E_{\hat{K}}^{(JM)} \right) 
\nonumber \\
& & \left( \sum_K A_{\hat{K}K}^{(JM)*} 
\Psi_{JMK}^* (\tilde{\Omega}^{t+1})
\right)
\label{propagator2} 
\end{eqnarray}

	The eigenfunctions of the symmetric top are (see Ref. \onlinecite{book_LevineMolecularSpectroscopy}):
\begin{equation} 
\Psi_{JMK}(\Omega)=
\left( \frac{2J+1}{8\pi^2} \right)^{1/2} 
\exp(iM\phi) d_{MK}^J (\theta ) \exp (iK\chi)
\end{equation}
where $d_{MK}^J(\theta) $ are the Wigner functions (given in Appendix B).
Using this expression we have:
\begin{equation}
\Psi_{JMK} (0)
= \left(\frac{2J+1}{8\pi^2}\right)^{1/2}
d_{MK}^J (0) = \left(\frac{2J+1}{8\pi^2}\right)^{1/2} \delta_{MK}
\label{theta0}
\end{equation}
and
\begin{equation}
\Psi_{JMK}^* (\tilde{\Omega})
= \left(\frac{2J+1}{8\pi^2}\right)^{1/2}
\exp( -iM\tilde{\phi}^{t+1}) d_{MK}^J (\tilde{\theta}^{t+1})
\exp( -iK\tilde{\chi}^{t+1})
\label{thetat}
\end{equation}
having made use of the following relation for Wigner functions:
$d_{MK}^{J}(0)=\delta_{MK}$.
M\"user and Berne concluded that, substituting these expressions,
Eq. \ref{propagator2} could be written as\cite{PRL_1996_77_002638}:
\begin{eqnarray}
& & \left\langle \omega^{t} \left\vert \exp \left( -\frac{\beta}{P} \hat{T}^{rot} \right) \right\vert \omega^{t+1}
\right\rangle  = \nonumber \\  
& = & \sum_J \sum_M \sum_{\hat{K}} 
\left( \frac{2J+1}{8\pi^2} \right) \left\vert A_{\hat{K}M}^{(JM)} \right\vert^2
\exp \left( -\frac{\beta}{P} E_{\hat{K}}^{(JM)}\right) d_{MM}^J (\tilde{\theta}^{t+1})
\exp( -iM(\tilde{\phi}^{t+1}+\tilde{\chi}^{t+1}))
\end{eqnarray}

	The propagator is a real quantity and this can be seen by symmetrising it with respect to $M$. 
This can be achieved by
calculating the average of the $M$ and $-M$ contributions for any given $J$ and $\hat{K}$:
\begin{eqnarray}
 \frac{1}{2}  
& & \left\{  
\left( \frac{2J+1}{8\pi^2} \right) \left\vert A_{\hat{K}M}^{(JM)} \right\vert^2 
\exp \left( -\frac{\beta}{P} E_{\hat{K}}^{(JM)} \right)
d_{MM}^J (\tilde{\theta}^{t+1}) \exp(-iM(\tilde{\phi}^{t+1}+\tilde{\chi}^{t+1})) 
\right.
\nonumber \\
& + & \left.  \left( \frac{2J+1}{8\pi^2} \right) \left\vert A_{\hat{K}\bar{M}}^{(J\bar{M})} \right\vert^2 
\exp \left( -\frac{\beta}{P} E_{\hat{K}}^{(J\bar{M})} \right)
d_{\bar{M}\bar{M}}^J (\tilde{\theta}^{t+1}) 
\exp(-i\bar{M}(\tilde{\phi}^{t+1}+\tilde{\chi}^{t+1})) \right\}
\label{eq_sum}
\end{eqnarray}
where $\bar{M}$ denotes $-M$. Given that (see Ref. \onlinecite{book_Zare_AngMom}):
\begin{equation}
d_{MM}^J (\theta) = d_{\bar{M}\bar{M}}^J (\theta)
\end{equation}
\begin{equation}
\left\vert A_{\hat{K} M}^{(JM)} \right\vert = \left\vert A_{\hat{K}\bar{M}}^{(J\bar{M})} \right\vert
\end{equation}
and
\begin{equation}
\exp \left( -\frac{\beta}{P} E_{\hat{K}}^{(JM)} \right) = 
\exp \left( -\frac{\beta}{P} E_{\hat{K}}^{(J\bar{M})} \right)
\end{equation}
Eq.\ref{eq_sum} can be simplified to:
\begin{equation}
\left( \frac{2J+1}{8\pi^2} \right)
\left\vert A_{\hat{K}M}^{(JM)}\right\vert^2 \exp \left( -\frac{\beta}{P} E_{\hat{K}}^{(JM)} \right)
d_{MM}^J (\tilde{\theta}_{t+1}) \cos(M(\tilde{\phi}_{t+1}+\tilde{\chi}_{t+1}))
\end{equation}
	Using this result, M\"user and Berne obtained the following final expression for the
orientational propagator:
\begin{equation}
\rho_{\mathrm{rot},i}^{t,t+1}(\beta /P)   = 
\sum_{J=0}^{\infty} \sum_{M=-J}^J \sum_{\hat{K}=-J}^J  
f_{i,J,M,\hat{K}}^{t,t+1} \exp \left( -\frac{\beta}{P} E_{\hat{K}}^{(JM)} \right)
\label{eq_rho_rot}
\end{equation}
where $f_{i,J,M, \hat{K}}^{t,t+1}$ is a function of the relative Euler angles between beads $t$ and $t+1$ and is given
by:
\begin{equation}
f_{i,J,M,\hat{K}}^{t,t+1} = 
\left( \frac{2J+1}{8\pi^2} \right) 
\left\vert A_{\hat{K}M}^{(JM)} \right\vert^2 
d^J_{MM} (\hat{\theta}_{i}^{t,t+1}) 
\cos[M(\hat{\phi}_i^{t,t+1}+\hat{\chi}_i^{t,t+1})] 
\end{equation}

{\bf C. Path integral simulations for a system of rigid rotors}

	Once the three contributions to the density function (Eq.\ref{eq_density_matrix}), 
the potential (Eq. \ref{eq_rho_pot}), the translational  (Eq. \ref{eq_rho_trans})
and rotational (Eq. \ref{equ_18} and \ref{eq_rho_rot}) 
 have been obtained, the partition function of a system of $N$ rigid rotors 
can be calculated by substituting them into  Eq. \ref{eq_Z_1}:
\begin{eqnarray}
\label{partition_function}
Z = Q_{NVT} & = & \frac{1}{N!}\left( \frac{MP}{2\pi\beta \hbar^{2} } \right)^{3NP/2} \int ... \int \prod_{i=1}^{N} \prod_{t=1}^{P}  d\mathbf{r}_i^t d{\bf \Omega}_i^t \\
& & \times \exp \left( - \frac{MP}{2\beta \hbar^2} \sum_{i=1}^N \sum_{t=1}^P\left( \mathbf{r}_i^t - \mathbf{r}_i^{t+1}\right)^2 - \frac{\beta}{P} \sum_{t=1}^P U^{t}\right) \nonumber \prod_{i=1}^N \prod_{t=1}^{P} \rho_{\mathrm{rot},i}^{t,t+1} (\beta /P),
\end{eqnarray}

If one uses  a pairwise potential then 
$U^t=\sum_i\sum_{j>i} U({\bf r}_i^t,{\bf \omega}_i^t,{\bf r}_j^t,{\bf \omega}_j^t)$.
One can now see that the partition function of a system of $N$ quantum
particles is isomorphic to that of a system of $N$ classical ring polymers.  
Each replica $t$ of molecule
$i$ interacts with the replicas with the same index
$t$ of the remaining particles through the
inter-molecular potential $U$,
and with replicas $t-1$ and $t+1$ of the same molecule $i$ through a harmonic
potential, whose coupling parameter depends on the mass of the molecules ($M$)
and on the temperature ($\beta=1/k_BT$),
and through  terms $\rho_{\mathrm{rot},i}^{t,t-1}$ and $\rho_{\mathrm{rot},i}^{t,t+1}$  which incorporate the quantisation
of the rotation and which depends on the relative orientations
of replicas $t-1$ and $t$, and $t$ and $t+1$. 
 The convention of the
Euler angles used in the present calculations, as well as 
the conversion from Cartesian coordinates to Euler angles, are
given in Appendix C. The procedure to obtain the 
Euler angles of replica $t+1$ in the body frame of replica $t$
is outlined in Appendix D.

Note that the function $\rho_{\mathrm{rot},i}^{t,t+1}(\beta /P)$ (Eqs. 38 and 39)  depends solely on two angles,
$\hat{\theta}^{t,t+1}$ and $\hat{\phi}^{t,t+1}+\hat{\chi}^{t,t+1}$.
It is  convenient to compute the density  function
for a grid of values over the angles $\hat{\theta} $ and $\hat{\phi} +\hat{\chi} $
and save these data in tabular form prior to any simulation. The value of
the density function for any particular $\hat{\theta} $ and $\hat{\phi} +\hat{\chi} $
can then estimated using an interpolation algorithm  in conjunction with
the tabulated data.

The internal energy \emph{E} can be calculated from the partition function $Z$ using 
the thermodynamic relation:
\begin{equation}
\label{Einterna}
E = -\frac{1}{Z} \frac{\partial Z}{\partial \beta}
\end{equation}
By substituting the partition function it can be shown that the internal
energy can be calculated as:
\begin{equation}
E = K_{tra} + K_{rot} + U,
\end{equation}
where the functional forms of these three components are:
\begin{equation}
\label{Ktra_eq}
K_{tra} = \frac{3NP}{2\beta} - \left\langle  \frac{MP}{2\beta^2\hbar^2} \sum_{i=1}^{N} \sum_{t=1}^{P} ({\bf r}_{i}^{t}-{\bf r}_{i}^{t+1})^{2} \right\rangle,
\end{equation}
\begin{equation}
\label{energy}
K_{rot} = \left\langle  \frac{1}{P} \sum_{i=1}^{N} \sum_{t=1}^{P} 
\frac{1}{\rho_{rot,i}^{t,t+1}} \left\{  \sum_{J=0}^{\infty} 
\sum_{M=-J}^{J} 
\sum_{\hat{K} =-J}^{J} f_{i,J,M, \hat{K}}^{t,t+1} 
E_{\tilde K}^{(J M)} e^{-\frac{\beta}{P}E_{\hat{K}}^{(JM)}} \right\} \right\rangle,
\end{equation}
\begin{equation}
\label{U_eq}
U = \left\langle \frac{1}{P} \sum_{t=1}^{P} U^t \right\rangle,
\end{equation}
        As with the rotational contribution to the density function, the numerator of the last term in Eq. \ref{energy}
was calculated prior to simulation for a grid of angles $\hat{\theta} $ and $\hat{\phi} +\hat{\chi}$
and saved in tabular form.

        The partition function for the $NpT$ ensemble can be calculated
using: 
\begin{equation}
Q_{NpT}= \int  \exp(-\beta pV) Q_{NVT} dV
\end{equation}
        The implementation of the $NpT$ ensemble in PIMC has
already been discussed in previous
works\cite{JCP_1989_90_5644,PRB_1995_51_2723}.

\section{Quantum effects in water}
\label{motivation}

	In this section we shall  see how the path integral formulation
for rigid rotors can be used to study quantum effects 
related to the atomic mass in water.
Water exhibits such quantum effects even at room temperature. For example,  
properties such as
the melting temperature,
the temperature of maximum density (TMD), or the heat capacity at
constant pressure, all show changes upon isotopic 
substitution\cite{book_FFranks_Water_Matrix_Life,chaplinweb}.
Experimental data for different water isotopes are given in
Table \ref{experiments}.
In particular, the melting temperature of water is about 4.49K higher 
in tritiated water than in H$_2$O. A more dramatic effect can be seen in the increase in the TMD, which is
about 8.91K higher in tritiated water.  
In contrast, isotopic substitution of the oxygen mass has little
influence on the properties of liquid water (see Table \ref{experiments}). 
This indicates that quantum effects are mainly due to the 
light mass of hydrogen, which leads to small values of the principal
moments of inertia. The lowest principal moment of inertia increases
when hydrogen mass increases, whereas it remains almost unchanged when
oxygen is substituted. The effect that this has on the magnitude of
quantum effects can be seen by using the following approximate expression
to estimate the quantum effects of a rigid asymmetric top\cite{AllenTildesleyBook}:
\begin{equation}
\label{eq_correction}
\frac{A-A_{cl}}{N}=\frac{\hbar^2}{24(k_BT)^2} \left[ \frac{\langle F^2 \rangle}{M}+
\frac{\langle \Gamma_A^2 \rangle}{I_A}+ \frac{\langle \Gamma_B^2 \rangle}{I_B} +
\frac{\langle \Gamma_C^2 \rangle}{I_C} \right] 
- \frac{\hbar^2}{24} \sum_{cyclic} \left(\frac{2}{I_A} - \frac{I_A}{I_BI_C} \right)
\end{equation}
where $F$ is the force that acts on the centre of mass of the particles,
$I_A$, $I_B$ and $I_C$ are the principal moments of inertia of the particles,
and $\Gamma_A$, $\Gamma_B$ and $\Gamma_C$ are the torques associated
with each principal axis of inertia.
This  expression indicates that both the low moments of inertia as well as 
the strength of the hydrogen bond (which leads to high values of
the average torque on the molecules) are responsible for 
the importance of quantum 
effects in water.

	The importance of nuclear quantum effects, as well as  the limitations of
performing classical simulations for water, have already been discussed by
Stillinger and Rahman in 1974,  one of the first articles published concerning simulations
of water\cite{JCP_1974_60_01545}. In this pioneering paper it was
pointed out that classical models are likely to overestimate 
the temperature difference between the TMD and the melting temperature
of water, which experimentally is 3.98K. They speculated
that this difference would be about 14K for classical models, which
was obtained by plotting the difference between the TMD
and the melting temperature for water, deuterated water and tritiated
water, as a function of the inverse of the hydrogen mass and then linearly
interpolating to infinite mass, which would correspond to the classical
limit. This hypothesis was also reiterated more recently by Guillot in his
review of  water models\cite{JML_2002_101_0219}.
A thorough investigation of classical rigid non-polarisable models eventually showed
that the temperature difference between the
TMD and the melting temperature was even larger; approximately 30K\cite{JCP_2005_123_144504}.

	The first simulations of water to   explicitly include quantum effects
were performed almost thirty years 
ago\cite{CPL_1984_103_0357,JCP_1985_82_05289,JCP_1985_82_05164,JCP_1991_95_03728}.
In these works, quantum effects in liquid water and deuterated water were investigated
by using the rigid non-polarisable ST2 model.
It was observed that liquid water is less structured than heavy water,  and that
classical water (which corresponds to the limit of infinite mass) 
is more structured than either of these isotopes.
In the same period, Wallqvist and Berne studied quantum effects 
using a flexible model\cite{CPL_1985_117_214}.

Since these seminal works, a good number of publications have appeared 
treating the quantum effects in water, the majority focusing on the liquid phase
\cite{JCP_1994_100_06692,JCP_2001_115_07622,JCP_2004_121_5992,JACS_2005_127_5246,JCP_2006_125_054512,JPCC_2008_112_00324}.
Some of these studies implemented  potentials designed to be used within classical
simulations. Even though this approach is valid when it comes to studying the effect that inclusion
of quantum effects have on the diverse properties of water, a quantitative description
requires the use of a potential that has been specifically designed to use within path
integral simulations. So far a few water models have been developed expressly to be used within quantum simulations.
These include both rigid\cite{JCP_2001_115_10758} and flexible models that have
either been fitted  to experimental data\cite{JCP_1997_106_02400,Manolopoulos_2009} or to more accurate \emph{ab initio}
calculations\cite{JCP_2008_128_074506}.
Recently more sophisticated path integral Car-Parrinello molecular simulations of
liquid water and ice I$_h$ have also been undertaken, which showed
that \emph{ab initio} calculations are also improved by incorporating nuclear
quantum effects\cite{PRL_2008_101_017801}.
As far as we are aware,  there are only a few studies that have considered quantum effects
for ice I$_h$\cite{JCP_1996_104_00680,JCP_2005_123_134502,PRE_2005_71_041204,JCP_2005_123_144506,JCP_2006_125_054512}.
There are also a couple of studies that have examined the liquid-solid\cite{JPCC_2008_112_00324,Manolopoulos_2009} 
and solid-vapour interface\cite{JPCC_2008_112_00324}.
\begin{table}
\centering
\caption{\label{experiments} Experimental data for different water isotopes.
The heat capacity at constant pressure is given at T=290K.}
\begin{tabular}{lccccc}
\hline \hline
 &  & $^1$H$_2$O   &  $^2$H$_2$O &  $^3$H$_2$O &  $^1$H$_2$$^{18}$O \\
\hline
T$_{melt}$ (K)                            &  & 273.15  &   276.97  & 277.64  &   273.43   \\
TMD   (K)                                 &  & 277.13  &   284.33  & 286.55  &   277.35   \\
$\rho_{TMD} \times 10^2$ (molec./{\AA}$^3$)& & 3.344   &   3.326   & 3.322   &  3.347 \\
TMD-T$_{melt}$ (K)            &  &   3.98    &     7.36  &   8.91  &     3.92   \\
C$_p$  (cal mol$^{-1}$ K$^{-1}$)&  & 17.93   &   19.60   & -     &     -  \\
\hline \hline
\end{tabular}
\end{table}

	In this work we treat the water molecule as being rigid. 
This means that our simulations are only capable of providing  information
about the low frequency inter-molecular librations (for
water these are below 900cm$^{-1}$), whereas 
high frequency intra-molecular vibrations (between 1500 cm$^{-1}$ 
for bending and 3500 cm$^{-1}$ for stretching)
will be ignored. 
The number of replicas required
to accurately reproduce the properties of a quantum system 
depends on the largest vibration frequency of the system:
\begin{equation}
P > \frac{\hbar \omega_{max}}{k_BT}
\end{equation}
From this expression it can be seen that at room temperature  approximately 
$P=$30 replicas should be used for
flexible models (for which the higher frequencies are around 3500 cm$^{-1}$), 
whereas $P=$5 or 6 is sufficient 
for a rigid model (for which the higher frequencies are about
1000 cm$^{-1}$).
This permits a considerable reduction in 
the computational cost of the simulations.

\section{Methodology}
\label{methodology}
	
	In the quantum simulations presented in this work, water was
described using the recently proposed TIP4PQ/2005 model\cite{Carl_2009,JCP_131_124518_2009},
which is the
quantum counterpart of the TIP4P/2005 model\cite{JCP_2005_123_234505}.
The classical TIP4P/2005 model was found to provide the best overall description of water from among
the many simple rigid non-polarisable models available in the literature\cite{FD_2009_141_0251,JCP_2006_125_034503,MP_2009_107_0365,JCP_2006_125_074510}.
In both models a 
Lennard-Jones (LJ) centre is located
on the oxygen site, positive charges on the hydrogens and a negative charge
along the bisector of the oxygen-hydrogen vectors. The total energy
of the system is given by:
\begin{equation}
U = \sum_i \sum_{j>i} \left\{  4\epsilon \left[ \left( \frac{\sigma}{r_{ij}} \right)^{12}
- \left( \frac{\sigma}{r_{ij}} \right)^{6} \right] + \sum_{m\in i} \sum_{n\in j} 
\frac{q_mq_n}{r_{mn}} \right\}
\end{equation}
where $r_{ij}$ represents the distance between the oxygen atoms in molecules $i$ and
$j$, $r_{mn}$ is the distance between the charge $q_m$ of molecule $i$ and
charge $q_n$ of molecule $j$. $\sigma $ and $\epsilon $ are the LJ
parameters. 
The parameters of the TIP4P/2005 and TIP4PQ/2005 models are given in Table \ref{tbl_model}. The only
difference between the two models is that the hydrogen charges
have been increased in the TIP4PQ/2005 model by 0.02$e$, which leads to 
stronger electrostatic interactions.
This
compensates for the loss of structure and the increase in energy that
is observed when quantum simulations are performed. The same recipe was
also used previously to obtain quantum counterparts of the classical 
SPC/F\cite{JCP_2006_125_184507} and TIP5P\cite{JCP_2001_115_10758} models.
The increase in the charges enhances the
dipole moment of the water molecule from 2.305D in TIP4P/2005 to 2.380D
in TIP4PQ/2005 (higher multipole moments will obviously also change accordingly). 
Note that in some cases for flexible models the change in geometry caused by the incorporation
of quantum effects with respect to the classical limit also leads to
an enhancement of the dipole moment of the water molecule. Therefore,
for flexible models it is generally not necessary to increase the charges in order to
perform quantum simulations\cite{Manolopoulos_2009}.

\begin{table}
\centering
\caption{\label{tbl_model} Parameters of the models TIP4P/2005 and TIP4PQ/2005.}
\begin{tabular}{lcccccc}
\hline \hline
   Model   &  $\sigma $({\AA }) &  $\epsilon /k_B$ (K) & $\angle$ HOH (deg) & $d_{OH}$  & $d_{OM}$ & $q_H$ (e) \\
\hline
TIP4P/2005                     & 3.1589  &   93.2  & 104.52  & 0.9572  &   0.1546 & 0.5564  \\
TIP4PQ/2005                    & 3.1589  &   93.2  & 104.52  & 0.9572  &   0.1546 & 0.5764  \\
\hline \hline
\end{tabular}
\end{table}

In this work, the influence of quantum effects in water was investigated
by performing $NpT$ PIMC simulations using the formulation for rigid rotors
proposed by M{\"u}ser and Berne\cite{PRL_1996_77_002638,JPCM_1999_11_0R117} described previously.
Classical $NpT$ simulations were also performed for comparison. The simulation
box contained 300 water molecules for the liquid phase, and 432 molecules
for ices I$_h$ and II. The initial proton disordered configuration of ice I$_h$ was 
obtained using the algorithm of Buch et al.\cite{JPCB_1998_102_08641,JCP_2004_121_10145}. 
The LJ interaction  was truncated at 8.5 {\AA}. 
Standard long-range corrections for the LJ part of the potential were added.
Coulombic interactions were calculated using Ewald summations.
Simulations usually consisted of about 30,000 cycles for equilibration
plus a further 100,000 cycles dedicated to obtaining averages. One MC cycle typically consisted
of $NP/2$ Monte Carlo moves, $N$ being the number of water molecules and $P$
the number of replicas of the system. The configurational space was explored by
using four types of movement attempts:
translation of one single bead of one molecule (30\%), rotation of a single bead
of one molecule (30\%), translation of a whole ring (20\%) and rotation of all the beads of a
given molecule (20\%). The maximum displacement or rotation was adjusted in
each case to obtain a 40\% acceptance probability. After each $NP/2$ Monte Carlo moves
one volume change was also attempted. The maximum volume change was adjusted
to obtain a 30\% acceptance probability.

\section{Results}

\begin{figure}[t!]
\begin{center}
\includegraphics[width=85mm,angle=0]{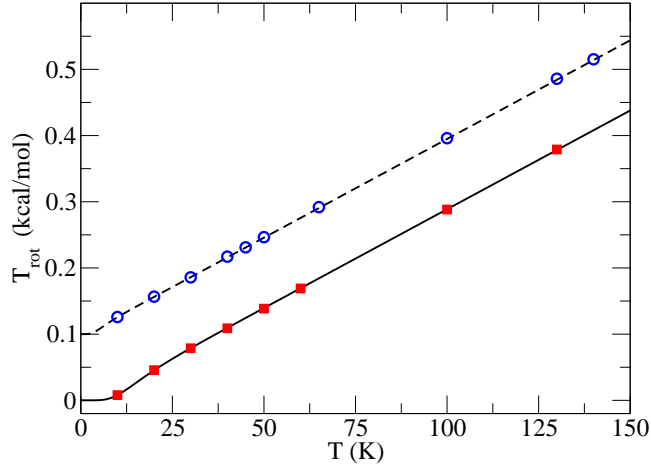}
\caption{\label{isolated_molecule} Kinetic rotational energy
of the isolated H$_2$O (filled red squares) and $^3$H$_2$O (open blue circles)
molecule as a function of temperature. 
There is a good agreement between the path integral simulations
and the rotational energy obtained from the theoretical
partition function of an asymmetric top for H$_2$O (solid line)
and $^3$H$_2$O (dashed line) geometry. For clarity, the rotational energy
of $^3$H$_2$O has been shifted 0.1kcal/mol in the $y$-axis.}
\end{center}
\end{figure}

Before presenting the results for bulk water, a preliminary  check was undertaken to show  that the rotational 
propagator was indeed able to reproduce the rotational kinetic energy
of a free asymmetric top, for which the rotational energy can be analytically
computed via evaluation of the partition function. The comparison between
the simulation data with the analytical expansion is excellent both for 
water and tritiated water (see Fig.~\ref{isolated_molecule}).
It can also be observed that for temperatures above approximately 50K
the quantum rotational energy of the free rotor is almost equal to
the classical value ($3/2k_BT$), which means that for an isolated rigid water molecule
quantum effects are only  important  below this temperature.
However, in condensed matter the situation is
different because there are inter-molecular forces, in this case hydrogen
bonds, that hinder the rotation of the molecules and lead to the 
appearance of noticeable quantum effects at much higher temperatures.

\begin{figure}[t!]
\begin{center}
\includegraphics[width=85mm,angle=0]{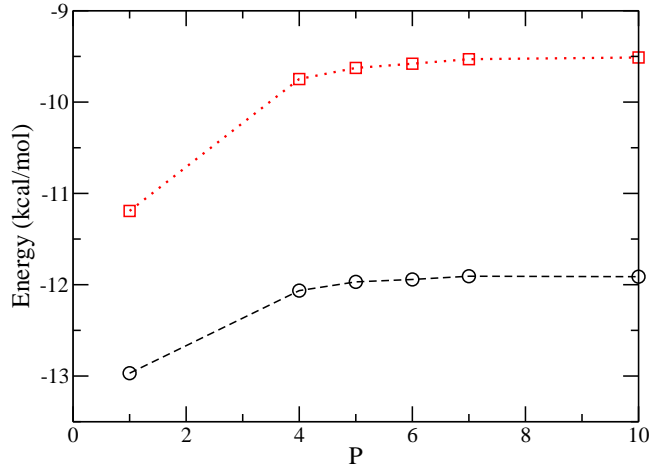}
\caption{\label{dependencia_p} Convergence of the potential energy $U$ (circles) and the
total energy $E$ (squares) as a function of the number of replicas $P$
for liquid TIP4PQ/2005 water at T=298K and $p=$1bar. The dotted and
dashed lines are only guides to the eye.}
\end{center}
\end{figure}

	Before performing simulations for the liquid and solid phases, we need
to choose the number of replicas, $P$, that are to be used in the simulations.
The classical limit corresponds to one single replica, whereas the quantum limit 
is approached as  the number of replicas tends to infinity. However, simulations
can only be performed for a finite number of replicas. In practice the number of replicas
is chosen so that it is small enough  that simulations do not become prohibitively
expensive, but high enough so as to capture the main contribution of the quantum effects.
$P$ represents a compromise between statistical convergence and theoretical
accuracy, so a study of how the desired property converges with
the number of replicas has to be carried out.
	We examined  the convergence of the potential energy and the total
energy as a function of the number of replicas for liquid water at $T$=298K
and $p$=1bar (see Fig.~\ref{dependencia_p}). It can be seen that both the
potential and total energies increase with the number of replicas. There is 
a large increase for small number of replicas and then  both magnitudes reach
a plateau above 5 or 6 replicas. The quantum limit would be obtained 
by plotting the total energy as a function of the inverse of the number
of replicas and taking the limit to infinite $P$. We found that
the total energy is lower than the value at $P\to\infty$ by about
3\% for $P=$5 and by about 2\% for $P=$7. In view of this we have chosen
to use $P=$5 replicas at room temperature. Other authors have also used a similar
number of replicas for water at room 
temperature\cite{JCP_1996_104_00680,JCP_2001_115_10758,JCP_2004_121_5992,JCP_2005_123_144506}.
For other temperatures, the number of replicas was chosen so as to keep
the product $PT$ approximately constant,
i.e. taking $P$=5 at $T$=300K we arrive at $PT\approx$ 1500K.

\subsection{Isotopic effects on the TMD and Cp}
\label{tmd}

	One of the idiosyncratic properties of water is the existence of a maximum
in density.
As mentioned previously, 
the location of the TMD is affected by variations in the hydrogen mass. In particular,
for deuterated water this maximum occurs 7K above the TMD of water and for
tritiated water this increases to  9K. 
Therefore, we expect that nuclear quantum effects will shift the TMD of water to lower
temperatures when quantum effects are implicitly incorporated.
In addition, given that good water models reproduce the 
TMD of water\cite{JCP_2004_120_09665,JCP_2005_123_234505}, it
would be interesting to check whether the TIP4PQ/2005 model is also able to
reproduce the experimental TMD. 

	The equations of state of water, deuterated water, tritiated water and 
classical water were calculated at $p=$1bar. $NpT$ PIMC simulations were performed
at six different temperatures in each case. As the maximum in density is
reflected in the third significant figure of the density, especially long simulations are
required to reduce the statistical error, so each simulation
consisted of at least 3 million MC cycles. The computational cost of the 
simulations was reduced by using the reaction field method\cite{steinhauser_mp}, rather than  
Ewald summations, to account for the long range electrostatic forces. 
It has been shown that Ewald summation and reaction field provide very similar results
for liquid water\cite{JCP_2002_117_08892} and for Stockmayer fluids\cite{CPL_1994_231_0366}. 
For the system sizes studied in this work ($N$=300-360)
the reaction field technique  yielded slightly higher energies and densities than the Ewald summations,
but the location of the TMD was unchanged\cite{JCP_131_124518_2009}.

\begin{figure}[t!]
\begin{center}
\includegraphics[width=75mm,angle=0]{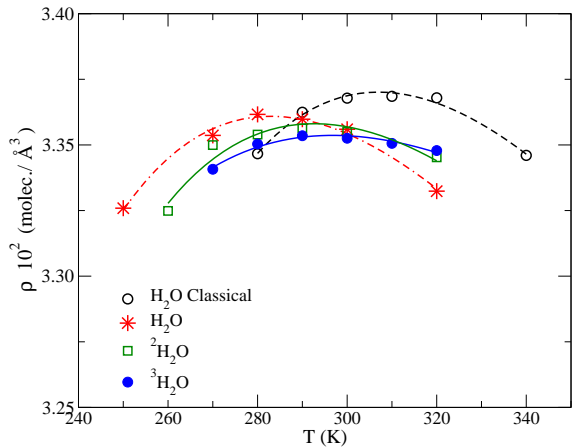}
\caption{\label{fig_tmd} Isotopic effects on the TMD of water along the room pressure isobar.
  Number densities (i.e number of molecules
  per unit of volume) as a function of temperature at room pressure
  are presented.}
\end{center}
\end{figure}

	The results are shown in Fig.\ref{fig_tmd}. The location
of the TMD was obtained by fitting the simulation data to a quadratic or cubic polynomial.
The TMDs obtained are given in Table \ref{tbl_tmd}. The results show that, using quantum simulations,
the TIP4PQ/2005 model 
predicts that the TMD of water occurs at 284(2)K, only 7K above the experimental result.
We have seen that upon increasing the number of replicas the TMD shifts to
280(2)K\cite{JCP_131_124518_2009}, indicating that  simulation results become even closer to
experimental results when more replicas are used.
	The results for deuterated water and tritiated water show that
TIP4PQ/2005 is also able to  qualitatively reproduce the shift to higher temperatures
when the mass of the hydrogen isotope increases, in line with experimental  observations. 
By considering the results for water, deuterated water and tritiated water (using  $P=5$ for 
the three molecules) one can see that  the location of the TMD shifts 8K for deuterated
water, and 12K for tritiated water, with respect to that of  water. 
These are only  slightly larger than the 7K and 9K, respectively, found experimentally.
	With regards to density, quantum effects affect differently the
density depending on whether we are above or below the TMD. For high temperatures
the number density increases with mass whereas for low temperatures the TMD the number density decreases
when the mass is increased (see Fig. \ref{fig_tmd}).

It is also interesting to study the shift in TMD when going from quantum
water to classical water. Our simulations predict that the TMD can
change as much as  30K when quantum effects are included. 
It seems that this is a typical shift 
for rigid non-polarisable
models (a similar result was found for the TIP5P model\cite{JCP_2001_115_10758}). Similarly
de la Pe\~na, Razul and Kusalik estimated a shift in the melting point of ice I$_h$ for the rigid 
TIP4P model of about
35K when nuclear quantum effects are included\cite{JCP_2005_123_144506}. 
However, the shift in the TMD between the quantum and classical limit
could well be different for different types of potentials (for example, flexible or polarisable models).
In particular, it has been found that for the flexible polarisable TTM2.1-F model\cite{JCP_2006_125_141102} the explicit inclusion of
quantum effects left the location of the TMD unchanged\cite{JCP_2007_127_074506}.
Further work is needed to clarify this. 
The TMD increases as the molecule becomes more and more classical (i.e. as  the mass of the hydrogen 
isotope is increased).  The number densities of water, deuterated water and 
tritiated water obtained from PI simulations along the room pressure isobar are shown in Fig. \ref{fig_tmd}. Results 
from classical simulations along this isobar are also presented. 
As can be seen in Fig. \ref{fig_tmd} the number density at the maximum is  hardly
affected by the mass of the hydrogen isotope;  
differences between the number densities at the maximum are 
within the estimated error bar. 
However, it seems that the number 
density at the maximum first decreases slightly 
on going from water to deuterated and then tritiated water, and then increases a little in the classical
limit. 
Experimentally it has been observed that the number density at 
the maximum decreases by $\approx$ 0.7\% on going from water to tritiated 
water (see Table \ref{experiments}). PIMC simulations, to a lesser extent, also reflect 
this decrease ($\approx$ 0.2\%).
We have shown in previous work 
that the internal energy of water is non-linear when plotted as a function of the inverse of the mass 
of the hydrogen isotope\cite{JCP_131_124518_2009}. The same also seems to be true here for the density of
water at the maximum.  This indicates that the behaviour of classical water cannot be obtained from a simple extrapolation 
of results obtained for water, deuterated water and tritiated water. 

	Another interesting calculation is the difference between 
the TMD and the melting
temperature obtained via classical and quantum simulations
($\Delta T=T_{TMD} - T_{melt}$). It has been found that for 
classical simulations of rigid non-polarisable models the TMD is situated
about 30K above the melting temperature
(i.e., $\Delta T$=30K)\cite{JCP_2005_123_144504},
which is much higher than the $\Delta T$=4K found
experimentally. 
Quantum simulations for the flexible q-TIP4P/F model predict that the
difference between the TMD and the melting point
is also about 30K\cite{Manolopoulos_2009}. Preliminary 
direct coexistence simulations\cite{ladd77,JCP_2006_124_144506}  of the melting temperature together
with the TMD calculations presented before 
indicate that the difference might be about 20-22K for the TIP4PQ/2005 model, which
improves upon the classical prediction, but that is still far from the 4K found
experimentally. This suggests that although the inclusion of 
nuclear quantum effects reduces the value of $\Delta T$
other features of real water need to be incorporated
in the model, such as  polarisability, 
in order to quantitatively reproduce  the experimental difference.
	
\begin{table}[!t]
\centering
\begin{tabular}{ccccc}
\hline\hline
System & & {TIP4PQ/2005}  & &  {Expt.}  \\
\hline
H$_2$O (P=5)          &    & 284(2) &  &   277.13 \\
$^2$H$_2$O (P=5)      &    & 292(2) &  &   284.33 \\
$^3$H$_2$O (P=5)      &    & 296(2) &  &   286.55 \\
Classical H$_2$O      &    & 307(2) &  &     --   \\
 \hline\hline
\end{tabular}
\caption{\label{tbl_tmd} Temperature of maximum density at
$p=$1bar as obtained from PIMC
simulations of the TIP4PQ/2005 model for several water isotopes (temperatures are given in kelvin).}
\end{table}

\begin{figure}[t!]
\begin{center}
\includegraphics[width=85mm,angle=0]{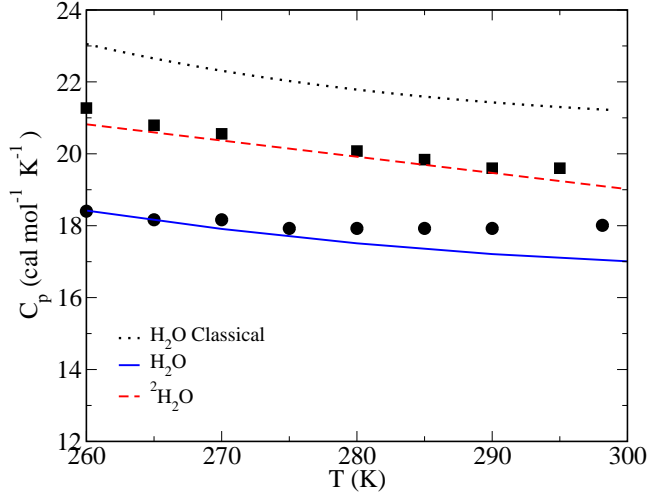}
\caption{\label{fig_cp} Heat capacity at constant pressure for water and other water isotopes
as a function of temperature at p=1bar calculated by means of PIMC simulations
with the TIP4PQ/2005. The heat capacity for classical water
was simulated with the TIP4P/2005. 
Experimental results for water\cite{JPCRD_2006_35_1021}
 (filled circles)
and deuterated water\cite{cp_h2o_d2O_exp} (filled squares) are also shown for comparison.}
\end{center}
\end{figure}

	From the simulations performed along the 1 bar isobar
it is straightforward to evaluate  
the heat capacity at constant pressure  for water as well as other
water isotopes
($C_p=\left.\frac{\partial H}{\partial T}\right\vert_p$, $H$ being the enthalpy).
It has been
found experimentally that the heat capacity of liquid water is considerably affected
by the isotopic substitution of the hydrogen atom (see Table \ref{experiments}). In particular,
the heat capacity is about a 10\% higher for deuterated water than for water
at room temperature\cite{JPC_1982_86_00998}.
The heat capacities obtained from PIMC simulations with TIP4PQ/2005 model for liquid water, deuterated water,
as well as  classical water simulated with the TIP4P/2005 model are shown in Fig.~\ref{fig_cp}.
It can be seen that quantum simulations with the TIP4PQ/2005 model are able to
quantitatively reproduce  the heat capacity of liquid water for a  broad 
range of temperatures. In addition, the increase in the heat capacity upon isotopic
substitution is also quite accurately captured. The agreement between simulations
and experimental data for deuterated water is quite remarkable.
Classical simulations with the TIP4PQ/2005 model 
are about 
25\% higher than those for quantum water using the 
same model\cite{JCP_2010_132_046101}, which again indicates
that $C_p$ is significantly affected by the inclusion of quantum effects.
One might think that a good description of the heat capacity 
of liquid water at room temperature could be obtained by using a
classical description with a model in which quantum effects are implicit
though the parametrisation of the model. However, to the best of our knowledge,
none of the rigid non-polarisable models proposed so far has been able to provide a
quantitative description of the heat capacity of 
liquid water\cite{JCP_2010_132_046101}. As an example, results for
TIP4P/2005 are also shown in Fig.\ref{fig_cp}, which are
in poor agreement both with experiments  and with
quantum simulations.
This suggests that quantum effects need to be incorporated explicitly in order to
obtain a quantitative description of the heat capacity of water.
This is further supported by the finding that PIMC simulations with TIP4PQ/2005
also reproduce the experimental heat capacity of ice I$_h$ from very low
temperatures up to room temperature (results for ice I$_h$ are shown later).
These results demonstrate that the main contribution can be captured using 
a rigid model and that the contribution from the intra-molecular degrees of
freedom is small. This is not unexpected; intra-molecular vibrations
exhibit very high frequencies ($\approx $ 3000 cm$^{-1}$) 
so at room temperature only the ground state is populated and
therefore there is little or no contribution to $C_p$ from the intra-molecular vibrations.

\subsection{Equation of state of ices}
\label{eos}

\begin{figure}[t!]
\begin{center}
\includegraphics[width=75mm,angle=0]{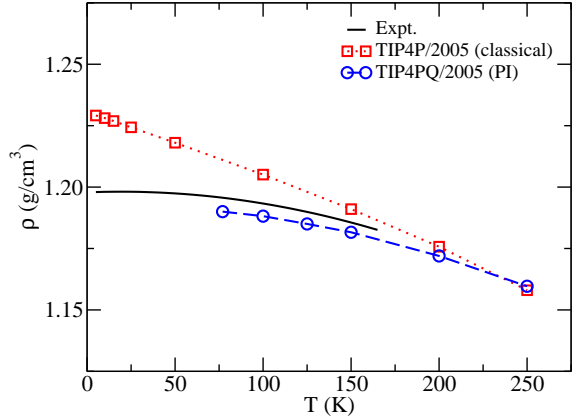}
\caption{\label{fig_eos} Equation of state of ice II at $p=$1bar
as calculated from PIMC simulations with TIP4PQ/2005 and from
classical MC simulations with TIP4P/2005. Experimental 
data are also shown for comparison\cite{JAC_2005_38_0612}.}
\end{center}
\end{figure}

	A significant deficiency of classical simulations is the inability
to reproduce the equation of state of solids at low temperatures. One of the consequences
of the third law of thermodynamics is that the thermal expansion coefficient
should tend to zero at zero temperature\cite{callen}, which is equivalent to saying that 
the density should remain constant at low temperatures. 
However, in previous work, we have seen that classical simulations of both
TIP4P/2005\cite{JCP_2005_123_234505} and TIP4P/Ice\cite{JCP_2005_122_234511} 
are unable to reproduce the correct curvature of 
the equation of state of ices at low temperatures\cite{JPCC_2007_111_15877}. 
This was attributed to the fact that quantum effects become increasingly
important as the temperature decreases.
To see whether the description of ice at low temperatures could be improved 
when quantum contributions were incorporated, we
performed PIMC simulations using the TIP4PQ/2005 model in order to obtain the equation of
state of ice II.
The results of quantum simulations with TIP4PQ/2005 
together with the results of classical simulations with TIP4P/2005 
and the experimental data of Fortes \emph{et al.}\cite{JAC_2005_38_0612}
are shown in Fig. \ref{fig_eos}. 
These results show that a good agreement
with the experimental data can be obtained 
when quantum contributions are explicitly included; 
the equation of state has the same curvature as 
the experimental curve, now in concordance with the third law of thermodynamics.
The same was also found to be true for ice I$_h$\cite{Carl_2009} and for 
hydrate sI\cite{JCP_2010_132_114503}.

\subsection{Structure of ices}
\label{structure}

\begin{figure}[t!]
\begin{center}
\includegraphics[width=75mm,angle=0]{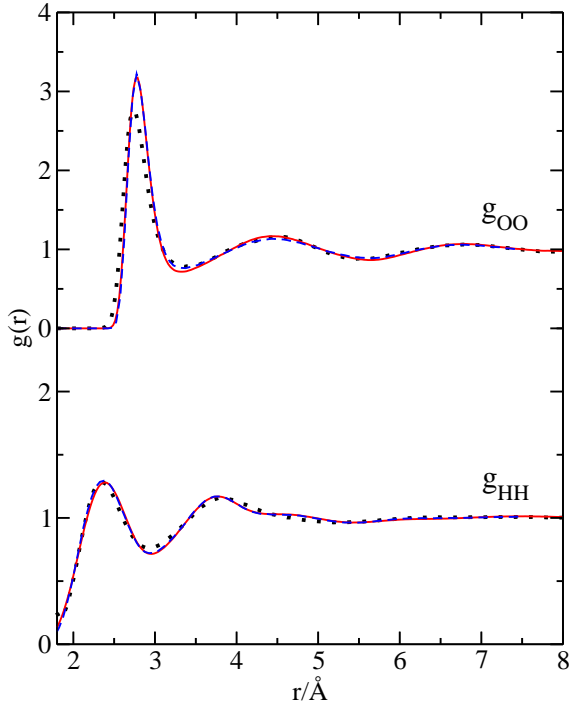}
\caption{\label{fig_rdf_liquid}Atomic distribution function of liquid water at 298K and 1 bar as calculated
from classical MC simulations with TIP4P/2005 (blue dashed line) and quantum PIMC
simulations with TIP4PQ/2005 (red solid line). Experimental data (black dotted line)
are also shown\cite{CP_2000_258_0121}.}
\end{center}
\end{figure}

\begin{figure}[t!]
\begin{center}
\includegraphics[width=75mm,angle=0]{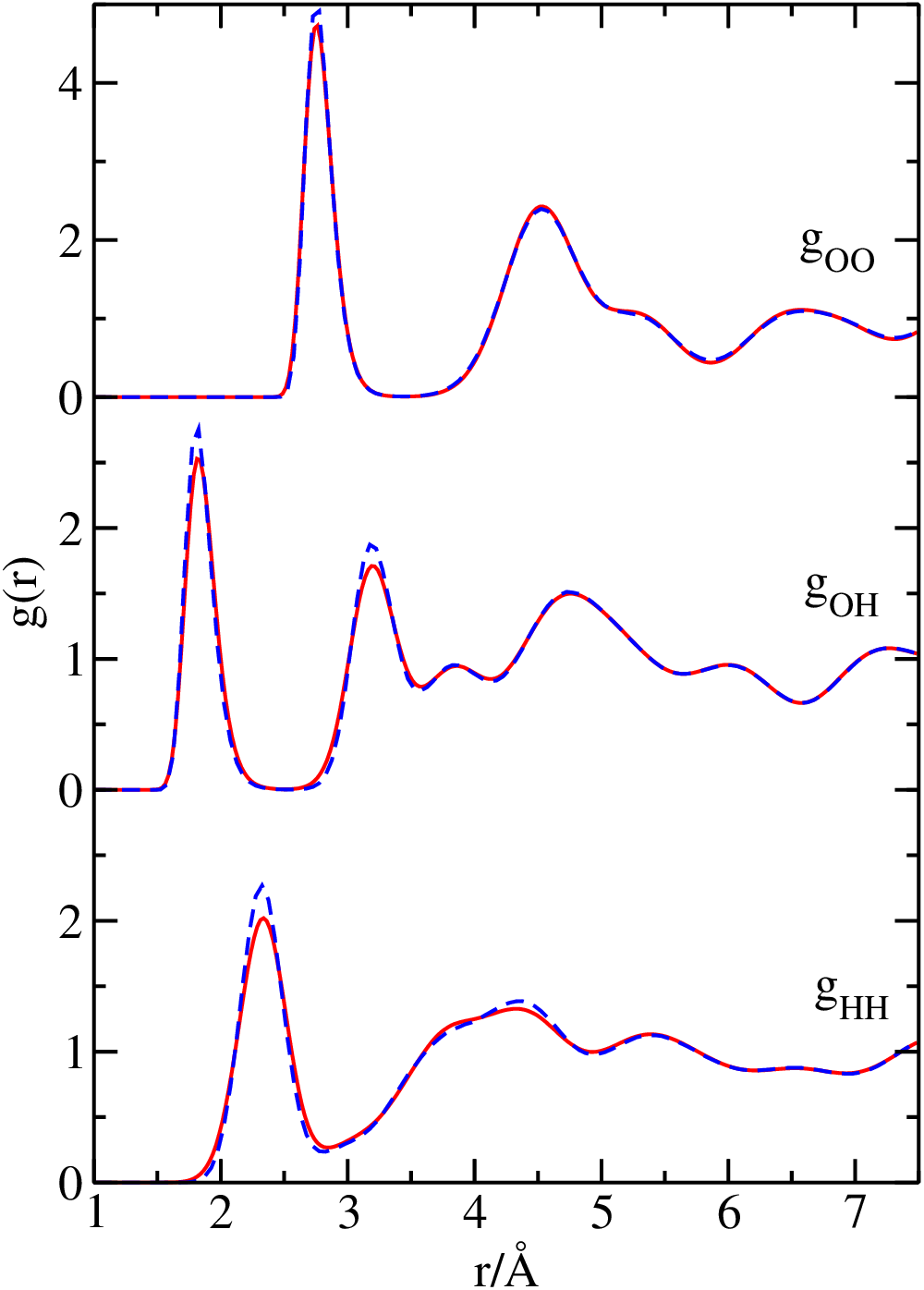}
\caption{\label{fig_rdf_Ih}Atomic distribution function of ice I$_h$ at 250K and 1 bar as calculated
from classical MC simulations with TIP4P/2005 (blue dashed line) and quantum PIMC
simulations with TIP4PQ/2005 (red solid line)}.
\end{center}
\end{figure}

\begin{figure}[t!]
\begin{center}
\includegraphics[width=75mm,angle=0]{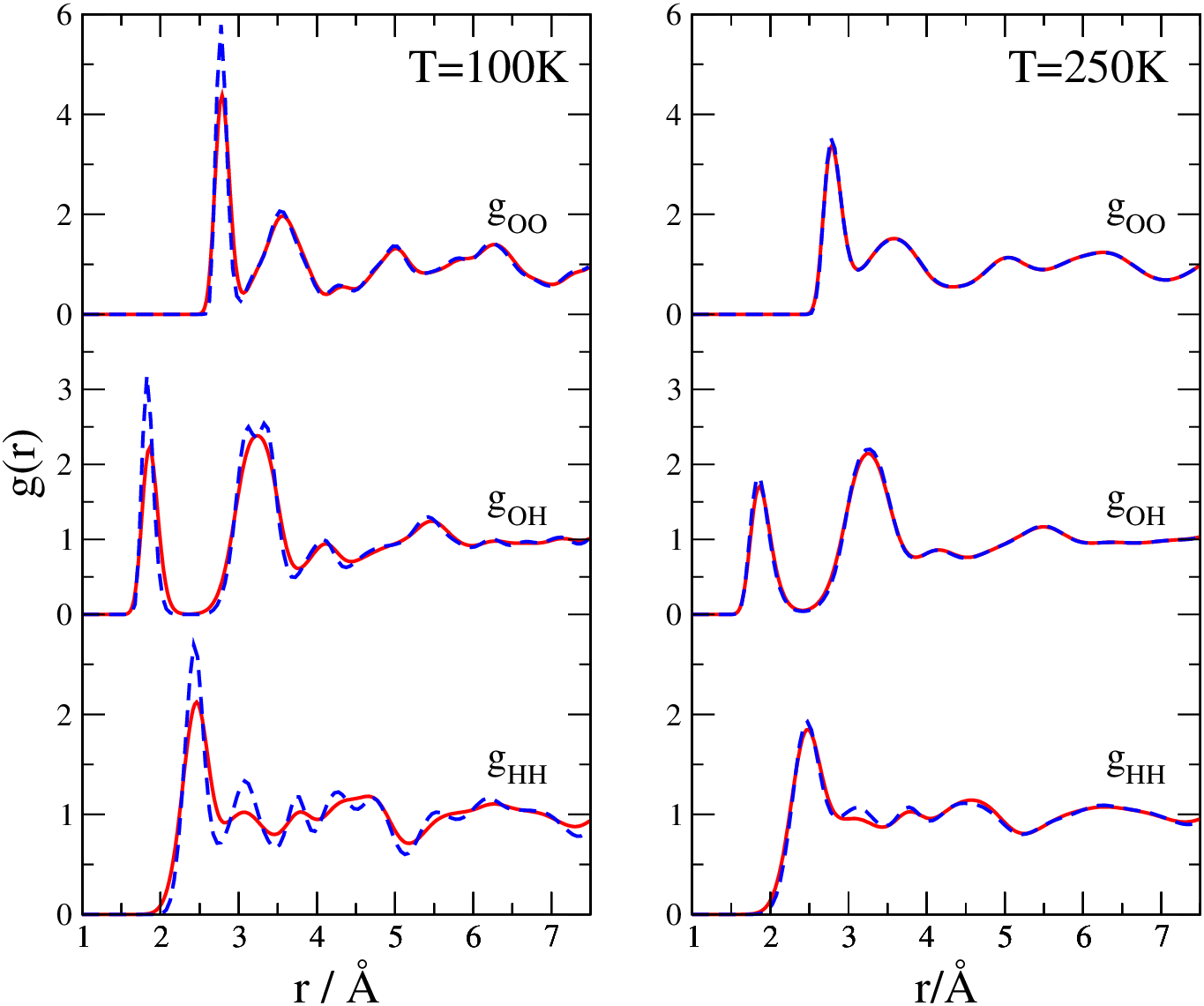}
\caption{\label{fig_rdf_II} Atomic distribution function of ice II at 100K and 250K and at 1 bar as calculated
from classical MC simulations with TIP4P/2005 (blue dashed line) and quantum PIMC simulations
with TIP4PQ/2005 (red solid line)}.
\end{center}
\end{figure}

	It is usually found that classical simulations using simple
models of water tend to overestimate the height of the first peak in the oxygen-oxygen
distribution function for both  liquid water\cite{JCP_2005_123_234505} and for ice I$_h$\cite{JPCC_2007_111_15877}.
It is well known that quantum effects lead to less structured
liquids\cite{JCP_1984_81_2523,JCP_1985_82_05289,MP_1993_78_1167,JCP_2004_121_5992} 
and solids\cite{JCP_1996_104_00680} and
so one might think that this can be corrected by including 
quantum effects. 
Radial distribution functions for liquid water and ices
I$_h$ and II obtained from classical simulations
for TIP4P/2005 and quantum simulations with TIP4PQ/2005 
at relatively high temperature (i.e. $T>$250K) and room pressure are
presented in Figs. \ref{fig_rdf_liquid}, \ref{fig_rdf_Ih} and \ref{fig_rdf_II}.
Differences are visible although relatively small, being larger 
for the oxygen-hydrogen and hydrogen-hydrogen distribution functions, which
is not unexpected since quantum effects are mainly due to  the hydrogen mass.
The situation is different at low temperatures. 
Both classical and quantum simulations were performed for ice II at $T=$100K.
The results are given in Fig. \ref{fig_rdf_II}. 
The first peak of the oxygen-oxygen
distribution function is considerably lower for quantum simulations with TIP4PQ/2005
than for classical simulations with TIP4P/2005. As far as we know, as yet there are no
experimental data for the atomic distribution function of ice II at this thermodynamic
state, but it is expected that quantum simulations provide a better
description of the structure at low temperatures (as was the
case for ice I$_h$\cite{Carl_2009}, the only ice for which the oxygen-oxygen atomic
distribution function has been experimentally measured).
The effects are much larger if one examines the oxygen-hydrogen and hydrogen-hydrogen
distribution functions, for which differences between classical and quantum simulations 
extend further than the first peak. In particular, 
the classical hydrogen-hydrogen distribution function exhibits a large number of
well defined peaks which become  considerably smoother in the quantum limit.
 The results for ice I$_h$ and II also suggest
that nuclear quantum effects might affect more
significantly the hydrogen-hydrogen atomic distribution function in
proton ordered ices, such  as ice II.

\subsection{Thermodynamic coefficients for ice I$_h$}
\label{coefficients}

	In previous work, we demonstrated  that classical simulations using both
 the TIP4P/2005 and the TIP4P/Ice models were unable to provide 
a good description for many thermodynamic coefficients for ice I$_h$\cite{JPCC_2007_111_15877}.
In particular, it was shown that classical simulations resulted in a
poor description of the heat capacity at constant pressure
and of the thermal expansion coefficient.
The thermal compressibility, on the other hand, was described reasonably well. 
We checked whether the description of some of these thermodynamic coefficients
could be improved by performing quantum PIMC simulations with the TIP4PQ/2005
model. The heat capacity at constant pressure and the thermal expansion
coefficient can be calculated from the simulations that trace 
out the room pressure isobar (data shown in Fig.\ref{fig_eos}).
The heat capacity was obtained by fitting the enthalpy to the
function $H=a+bT^2+cT^3$ and differentiating this fit with respect 
to the temperature. 
The isothermal compressibility was calculated by 
performing simulations at $p=$-500, -250, 0, 250 and 500 bars for temperatures
between 100K and 250K.
The density along each of these isotherms could be nicely fitted to a straight line.
The isothermal compressibility was computed by differentiating the
density with respect to the pressure from the fit
($\kappa_T=\frac{1}{\rho}\left.\frac{\partial \rho}{\partial p}\right\vert_T$).
Once the thermal expansion coefficient and the isothermal compressibility
are known, the pressure coefficient ($\beta_V=\left.\frac{\partial p}{\partial T}\right\vert_V$) can be readily computed  via $\beta_V=\alpha /\kappa_T$.

\begin{figure}[t!]
\begin{center}
\includegraphics[width=115mm,angle=0]{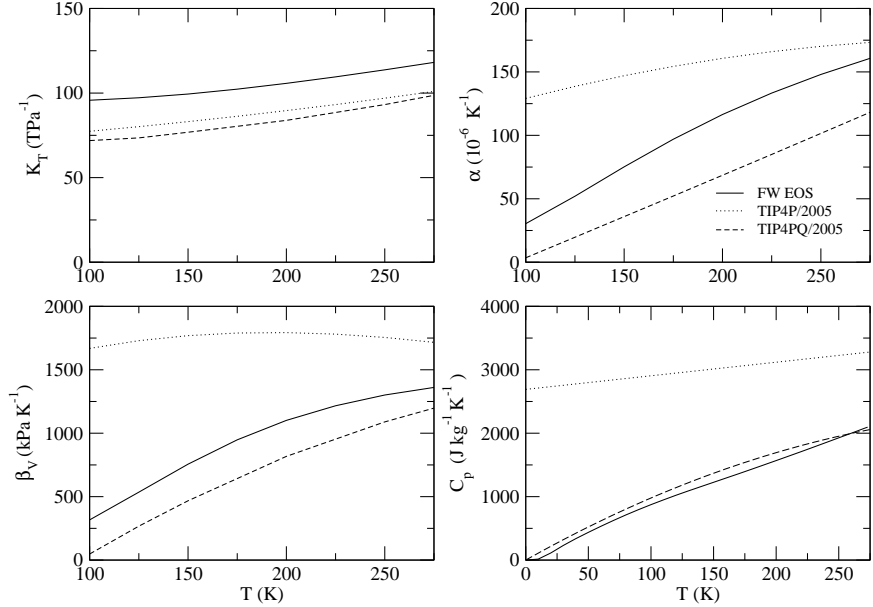}
\caption{\label{fig_coefficients} Thermodynamic coefficients ($\kappa_T$, $\alpha$, $\beta_V$ and C$_p$) of ice I$_h$ at 
p=1bar, as calculated from classical simulations with TIP4P/2005 and from
quantum simulations with TIP4PQ/2005. Experimental data are also shown
for comparison\cite{JPCRD_2006_35_1021}.}
\end{center}
\end{figure}

	The thermal coefficients as obtained from PIMC simulations using the
TIP4PQ/2005 model, as well as those form classical MC simulations using the
TIP4P/2005 model \cite{JPCC_2007_111_15877} are shown in Fig. \ref{fig_coefficients}. The thermal coefficients
derived from the experimental equation of state of Feistel and Wagner are also
shown for comparison\cite{JPCRD_2006_35_1021}. The results indicate that,
except for the thermal compressibility, for which a similar accuracy is obtained
in quantum and classical simulations, quantum simulations considerably
improve the description of the thermodynamic coefficients. In particular,
the heat capacity at constant pressure, which was not reproduced by classical 
simulations at any temperature, is now nicely reproduced from room temperature
all the way down to zero kelvin. This is in line with the results for liquid water, where again
the experimental heat capacity could only be reproduced by the
explicit inclusion of quantum effects. These results strongly suggest that 
quantum effects are crucial when it comes to describing the heat capacity of either liquid
water or ice I$_h$. 

	Regarding  the coefficient of thermal
expansion $\alpha$, quantum simulations using the TIP4PQ/2005 model 
give a much better description than classical simulations with the TIP4P/2005 model.
Even though at room temperature classical simulations predict a value closer
to the experimental data, quantum simulations provide a better overall description
over the whole range of temperatures. Importantly, the thermal expansion coefficient
now tends to zero at zero kelvin, as it should according to the third law of
thermodynamics. Finally, the description of the pressure coefficient ($\beta_V$) is also considerably
improved when including quantum effects. This simply reflects  that the thermal
expansion coefficient is improved in a quantum description of the system, 
since $\beta_V=\alpha /\kappa_T$.
The isothermal compressibility
is little affected by quantum effects; almost all the change in the
pressure coefficient is due to    a good description of the thermal expansion
coefficient.

\subsection{Relative energies of ices at zero kelvin}
\label{relative_energies}
 
	Finally we have also computed the relative energies between various ice phases 
at zero temperature. It has been found that some classical water models
result in a rather good description of the phase diagram of 
water\cite{PRL_2004_92_255701,JCP_2005_122_234511,JCP_2005_123_234505}.
However, there is still room for improvement. For example, it has been found that 
usually ice II is over-stabilised with respect to ice I$_h$, for some models, so much so
that ice II   completely removes ice I$_h$ from the phase diagram\cite{JCP_2007_127_154518}.
A preliminary outline of the phase diagram for a particular model 
can be obtained by estimating the coexistence pressures between the competing solid
phases at zero temperature. At zero kelvin phase transitions occur
with zero enthalpy change, so a calculation of entropy is avoided.
Assuming  that the change in energy and density between two solid
phases is almost independent of pressure at zero kelvin (which is indeed 
a rather good approximation for ices), the calculation of coexistence pressures
between two ices at zero temperature can be estimated from\cite{JCP_1984_81_04087}:
\begin{equation}
p=\left. -\frac{\Delta E }{\Delta V} \right\vert_{p=0}
\end{equation}
	Therefore, by simply calculating the energy and density of the solid
phases at zero temperature and zero pressure one can obtain a reasonable
estimate of the coexistence pressure at zero temperature.

\begin{figure}[t!]
\begin{center}
\includegraphics[width=80mm,angle=0]{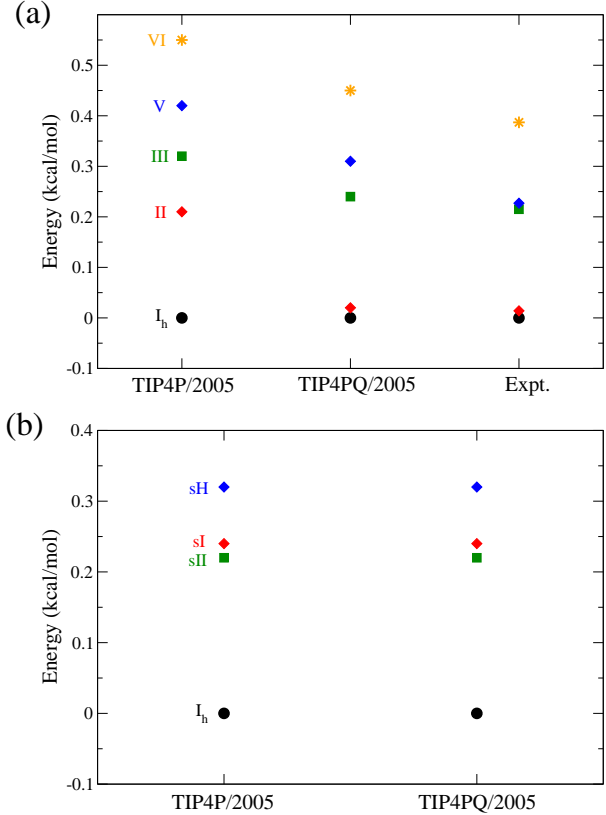}
\caption{\label{fig_relative_energies} 
Estimate of the relative energies
at zero temperature of the ice phases as calculated from classical MC
simulations with TIP4P/2005 and quantum simulations with TIP4PQ/2005
model. These data are taken from 
Refs. \onlinecite{JCP_2007_127_154518,JCP_2009_131_034510,Carl_2009,JCP_2010_132_114503}.
Experimental data taken from Ref. \onlinecite{JCP_1984_81_04087} is also
given for comparison.}
\end{center}
\end{figure}

	The properties at zero temperature were computed for ices
I$_h$, II, III, V and VI. Empty hydrates structures sI, sII
and sH, which have been shown to be the stable solid  phases at negative
pressures\cite{JCP_2009_131_034510}, were also considered. Simulations 
were performed along the zero bar isobar in the temperature range  
from 250K to 77-100K. The energy at zero temperature was obtained
by fitting the data to the  function $E=a+bT^2+cT^3$, from which one can estimate $E(T=0K)$. 
The energies obtained using this procedure are represented in 
Fig. \ref{fig_relative_energies}. Energies are given relative
to the energy of ice I$_h$, which experimentally is the most stable phase
at zero temperature and at zero pressure.
The results show that both classical MC simulations using TIP4P/2005
and quantum PIMC simulations using TIP4PQ/2005 predict
that ice I$_h$ is the most stable phase, in agreement with experimental results.
It is also observed that the relative energies of ices II, III, V and
VI change when quantum effects are explicitly taken into account.
In particular, ice I$_h$ is destabilised with respect to ice II by about 0.2kcal/mol,
so that now the relative stability of ice II with respect to ice
I$_h$ is much closer to the experimental value. This indicates that quantum effects
are larger in ice I$_h$ than in ice II, resulting in a de-stabilisation
of the former. Ices III, V, and VI are also stabilised with respect to ice
I$_h$, but to a smaller extent (by about 0.1 kcal/mol). Finally, the relative
stability of the empty hydrate structures are not appreciably
changed. 
Taking everything into account, we can identify three different
families of ices according to the importance of quantum effects.
The first family includes ice I$_h$ and the empty hydrate structures sI, sII
and sII, which are influenced the most by quantum effects. The second
family will be that formed for ices III, V, and VI, and finally, ice II, which
is the least affected by quantum effects, forms the third family. 

	The reason why quantum effects make distinct contributions  to 
the various ice phases can be understood by looking at the geometrical
arrangement of the four molecules that form a hydrogen bond
with a central one. These molecules form a nearly perfect tetrahedron
in ice I$_h$ and a slightly deformed tetrahedron with deviations of about
10 degrees in the hydrates. On the other hand the deviations from
the perfect tetrahedron are of about 30 degrees for the ices II, III,
V and VI. As a result, the strength of the hydrogen bond is
larger in ice I$_h$ and the hydrates than in the remaining ices and, 
as can be seen in Eq. \ref{eq_correction}, this results in an increased value for the average forces and torques,
boosting the quantum influences. 

	Once the energies at zero temperature have been calculated,
the coexistence pressures can also be obtained. The results are given 
in Table \ref{tbl_coexistence_pressure}. As can be seen, the coexistence
pressure between ices I$_h$ and II is the most affected coexistence line. 
It decreases from about 2090 bar in classical simulations
(TIP4P/2005) to about 195 bar in quantum simulations (TIP4PQ/2005),
resulting in a much better agreement with the experimental results.
The coexistence lines between the remaining ices (II, III, V
and VI) are also affected, but to a lesser extent. In general, 
quantum results are closer to the experimental data than the
results obtained using classical simulations.
Finally, coexistence pressures between I$_h$ and the hydrates
or between hydrates themselves are largely unaffected by quantum effects.

In summary, phase transitions between solid
phases belonging to the different families are the most affected by the
inclusion of quantum effects, whereas phase transitions involving
phases of the same family are in general less affected.
Therefore, the explicit inclusion of quantum effects is crucial if
one wishes to reproduce the phase transitions between solid phases
belonging to different ice families, especially the transition I$_h$-II,
whereas usually phase transitions between ice phases of the same
family can be calculated by means of classical simulations in conjunction with a
good classical model.

\begin{table}[!t]
\centering
\begin{tabular}{ccccccc}
\hline\hline
Phase & & TIP4P/2005  &  & TIP4PQ/2005 & &  Expt.  \\
\hline
I$_h$-II   & &  2090 & &   195 & &   140(200)  \\
I$_h$-III  & &  3630 & &  2727 & &  2400(100)  \\
II-V       & & 11230 & & 15731 & & 18500(4000)  \\
II-VI      & &  8530 & & 10935 & & 10500(1000) \\
III-V      & &  3060 & &  1998 & &  3000(100)  \\
V-VI       & &  6210 & &  6848 & &  6200(200)  \\
I$_h$-sI      & & -4174 & & -3948 & & --     \\
I$_h$-sII     & & -3379 & & -3249 & & --     \\
I$_h$-sH      & & -4072 & & -3933 & & --     \\
sII-sI     & &  2787 & &  2267 & & --     \\
sII-sH     & & -7775 & & -7557 & & --     \\
 \hline\hline
\end{tabular}
\caption{\label{tbl_coexistence_pressure} Coexistence pressures (in bar) at
zero temperature obtained from quantum PIMC simulations with
TIP4PQ/2005 and classical MC simulations with TIP4P/2005.
These data are taken from Refs. \onlinecite{JCP_2007_127_154518,JCP_2009_131_034510,Carl_2009,JCP_2010_132_114503}.
Experimental data taken from Ref. \onlinecite{JCP_1984_81_04087} is also
given for comparison.}
\end{table}

\section{Conclusions}
\label{conclusions}

	In this work it has been  shown in detail how the formulation
of the path integral for rigid rotors, derived by M{\"u}ser and Berne\cite{PRL_1996_77_002638},
can be applied to water. Using this formulation, a large number
of properties of liquid water, ices and hydrates have been studied
by PIMC simulations using the TIP4PQ/2005 model, which was specifically designed
to use within quantum simulations\cite{Carl_2009}.
For liquid water, isotopic effects on the TMD and on the heat capacity
have also been considered. 
The results show that in general a better description of water is
obtained when quantum effects are included, although some properties can
also be reasonably described in classical simulations with a good classical
model. In addition, PIMC simulations with TIP4PQ/2005 reproduce
the experimental isotopic effects on the TMD and the heat capacity of
liquid water.

Quantum effects have been found to be crucial when it comes to reproducing many
properties of water and ices. In particular, quantum effects
have been found to be most important with regards to  the properties
of ices at low temperatures, which is not entirely unexpected.
In particular, classical simulations fail to reproduce the curvature of the
equation of state at low temperatures found experimentally and
imposed by the third law of thermodynamics\cite{JPCC_2007_111_15877}. Our results
show that this can be corrected by including quantum effects
resulting in physically agreeable equations of state for ices I$_h$\cite{Carl_2009} and II 
over a quite broad range of temperatures.
Classical simulations overestimate the first peak in
the oxygen-oxygen radial distribution function of ice I$_h$ at 77K, which
again is brought into to agreement with experiment when quantum
simulations are performed\cite{Carl_2009}. The same behaviour was found for ice
II and is expected to occur for other ices. 
As a result of the better description of the equation of state
the coefficient of thermal expansion
at low temperatures of ice I$_h$ is also improved when quantum effects are
included.
In addition to the properties of ices at low temperatures,
quantum effects have also been found to be important
when it comes to  reproducing the heat capacity of ice I$_h$ and water
at all temperatures.
In addition, isotopic effects on the heat capacity of liquid
water have also been captured. 

	We also found that the magnitude of quantum effects
is different for different ices and, therefore, they need 
to be included if one wishes to improve the description of phase transitions.
In particular, it has been found that ices can be classified
into three different families, according to the importance of
quantum effects: the first family is formed by ice I$_h$ and the
hydrates structures sI, sII and sH, for which quantum effects
are the largest, the second family comprises ices III, V and VI,
and the third family is formed by ice II, for which
quantum effects are the smallest. Phase transitions between ices
belonging to different families change when quantum effects
are included, whereas transitions between ices belonging to the
same family are only slightly affected by quantum effects.
As quantum effects are also different for liquid water and
ice I$_h$, the melting point of water is also affected by
quantum effects; it has been found in previous works
that the melting point shifts to lower temperatures when
quantum effects are included\cite{Manolopoulos_2009,JCP_2005_123_144506}.
	In general, quantum affects should also
affect any property that involves two phases for which
quantum effects are different. For example, in previous
work it has been shown that quantum effects improve the
description of the enthalpy of vaporisation\cite{JCP_131_124518_2009} and the sublimation
enthalpy\cite{JCP_2010_132_114503}.

	However, it has been found that other room temperature properties,
although also affected by quantum effects, can be properly
described using a classical model.
This can be explained because classical models are usually fitted
to reproduce some experimental data at room temperature, so
in some way quantum effects at this temperature are implicit in the model.
For example, it has been found that the structure of liquid
water and ices I$_h$ and II above 250K is reproduced with similar
accuracy in classical and quantum simulations (although only
quantum simulations can reproduce isotopic effects)\cite{JCP_131_124518_2009}.
The densities of ices can also be reproduced at room temperature
with good accuracy using classical simulations\cite{Carl_2009}.
The isothermal compressibility of ice I$_h$ shows little improvement  
with respect to classical simulations at all temperatures, which
also might be related to the fact that quantum effects are 
influenced little  by pressure.

With regards to the TMD, several classical models were proposed
that reproduce the experimental TMD\cite{JCP_2005_123_234505,JCP_2004_120_09665}, although the temperature
difference between the TMD and the melting temperature is
largely overestimated (they usually predict a 30K difference\cite{JCP_2005_123_144504}
instead of the 4K found experimentally). Preliminary 
PIMC calculations indicate that this difference might be reduced
for TIP4PQ/2005 to $\approx22$K, which means that, although
some improvement is achieved, other features of 
real water need to be included, such as polarisability and
flexibility, in order to obtain a better agreement with
experiment. 

By using a rigid model, we have ignored the influence of quantum effects
in the intra-molecular degrees of freedom. 
Despite this seemingly drastic approximation the results presented here seem to indicate that for many properties
the main quantum contributions arise from the inter-molecular
degrees of freedom.
Comparison between
quantum simulations of rigid and flexible models will be 
be very useful to quantify the relative importance of quantum effects
on inter-molecular and intra-molecular degrees of freedom.

\acknowledgements

This work was funded by grants FIS2010-16159, FIS2010-15502 and FIS2009-12721-C04-04 
of Direcci\'on
General de Investigaci\'on, S2009/ESP-1691 of
Comunidad Aut\'onoma de Madrid and 200980I099 of CSIC.
Useful discussions with Prof. J.L.F. Abascal are gratefully acknowledged.

\newpage

\appendix
{\bf Appendix A. Asymmetric top eigenfunctions}
	
\begin{figure}[t!]
\begin{center}
\includegraphics[width=60mm,angle=0]{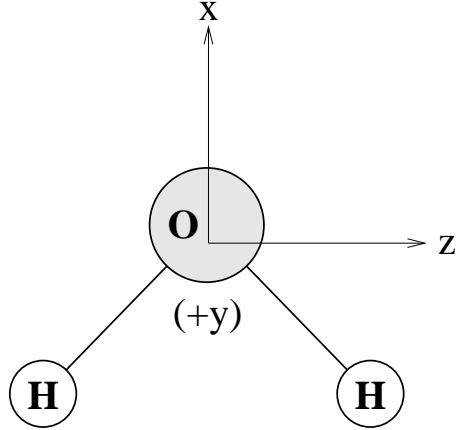}
\caption{\label{fig_axis} Reference system fixed in the water molecule
adopted in this work. This convention is usually denoted as the
$bca$ convention, because the $x$,$y$ and $z$ axis coincide with the
$b$, $c$ and $a$ principal axes of inertia (thus
$I_a<I_b<I_c$). The origin of the coordinate system
is located at the centre of mass of the water molecule.}
\end{center}
\end{figure}

	The rotation of an asymmetric free rotor can be obtained by solving the
Schr\"odinger equation:
\begin{equation}
\hat{H}^{rot} \Psi =  \hat{T}^{rot} \Psi = E \Psi
\end{equation}
where $\hat{T}^{rot}$ is the Hamiltonian associated with the angular momentum:
\begin{equation}
 \hat{T}^{rot} =\frac{ \hat{L}_x^2 }{ 2 I_{xx}} + \frac{\hat{L}_y^2}{2 I_{yy}} 
+ \frac{\hat{L}_z^2}{2 I_{zz}}
\end{equation}
In this equation $\hat{L}_x$, $\hat{L}_y$ and $\hat{L}_z$ are the three components of
the angular momentum and $I_{xx}$, $I_{yy}$ and $I_{zz}$ are the 
three components of the momentum of inertia. 
To solve this equation it is convenient to choose a reference system
so that the $x$, $y$ and $z$ axis are located along the three
principal axes of inertia, denoted as $a$, $b$ and $c$.
We adopt the convention that $I_a \leq I_b \leq I_c$. 
Note that there is no unique way to identify 
$x, y, z$ with $a, b, c$ (see Refs. \onlinecite{book_Zare_AngMom,bunker}). 
For example, one could associate $x$ with $a$, $y$ with $b$, and $z$ with $c$, which
is usually referred to as $abc$ convention. Alternatively we
could choose to identify $x$ with $b$, $y$ with $c$ and $z$ with $a$,
which is usually known as $bca$ convention, and it is this convention
that was used in this work (see Fig. \ref{fig_axis}). The choice of  axis is highly relevant since
it defines the Euler angles that appear in the three components of
the angular momentum.

	In an asymmetric top all three moments of inertia are 
distinct ($I_a\neq I_b\neq I_c$). In this situation the Hamiltonian
commutes with $\hat{L}^2$ and with $\hat{L}_z$, but not with $\hat{L}_c$: 
\begin{eqnarray}
[ \hat{L}^2 ,  \hat{H}^{rot} ] =   0  \hspace*{1cm} 
[ \hat{L}_z ,  \hat{H}^{rot} ] =   0  \hspace*{1cm}
[ \hat{L}_c ,  \hat{H}^{rot} ] \neq  0 
\end{eqnarray}
	Therefore, the eigenfunctions of the Hamiltonian will also
be eigenfunctions of $\hat{L}^2$ and $\hat{L}_z$, but not of $\hat{L}_c$:
\begin{eqnarray}
\hat{L}^2 \Psi & = & J (J+1) \hbar^2 \Psi \hspace*{1cm} J= 0, 1, ..., \infty \\
\hat{L}_z \Psi & = & M \hbar \Psi \hspace*{1cm} M=-J, ..., 0, ..., J
\end{eqnarray}
	The solutions for the Schr\"odinger equation for an asymmetric top
will be denoted as $| J M \hat{K} \rangle $. 
The integer $\hat{K}$ is not a true quantum number (i.e., it does not
quantise any observable) it is simply a number used to label the
$(2J+1)$ possible values of the energy available for each value
of $J$ and $M$.
The functions
$| J M \hat{K} \rangle $ can
be obtained by expanding them in a basis set formed by the
eigenfunctions of the spherical top
($| J M K \rangle $):
\begin{equation}
| J M \hat{K} \rangle = \sum_{K} A_{\hat{K}K}^{JM} | J M K \rangle
\end{equation}
	Using the $bca$ convention, the energies $E_{\hat{K}}^{(JM)}$ 
of the asymmetric top and 
the coefficients $A_{\hat{K}K}^{(JM)}$ (i.e, the eigenvectors 
$A_{\hat{K}}^{(JM)}$)  can be obtained solving 
the following secular determinant (see Refs. \onlinecite{zare,bunker})
for each value of $J$ and $M$ :
\begin{eqnarray}
H_{KK} & = & \frac{1}{2} (B+C) [ J(J+1) - K^2] + A K^2 \nonumber \\
H_{KK\pm 2} & = & \frac{1}{4} (B-C) [J(J+1) - K(K\pm 1)]^{1/2}
                             [J(J+1) - (K\pm 1)(K\pm 2)]^{1/2}
\end{eqnarray}
	where $K$ ranges from $-J$ to $+J$. The remaining  elements of
the determinant are zero.
$A$, $B$ and $C$ are the rotational constants, 
$A=\frac{\hbar}{4 \pi I_{a}}$,
$B=\frac{\hbar}{4 \pi I_{b}}$ and $C=\frac{\hbar}{4 \pi I_{c}}$. Note that
since $I_a \leq I_b \leq I_c$, it follows that $A \geq B \geq C$.
This determinant has dimensions of $(2\times J + 1)$.
Therefore, $(2\times J + 1)$ eigenvalues are obtained for each
value of $J$ and $M$, which are, in general, all different. These
$(2\times J + 1)$ energy levels are labelled with the $\hat{K}$ index.
However, as $M$ does not appear in the
determinant, there is a $(2\times J+1)$ degeneracy in the energy
associated with $M$.
Note that if we had chosen a different convention for the reference
system, e.g. the $abc$, the secular equation that we would need to solve
would be different (see, e.g., Refs. \onlinecite{zare,bunker,book_LevineMolecularSpectroscopy}).


{\bf Appendix B. Wigner functions} \\

	The Wigner functions are given by\cite{GrayGubbins}:
\begin{eqnarray}
d_{MK}^J (\theta) &=& 
[ (J+M)! (J-M)! (J+K)! (J-K)!]^{1/2} \\  \nonumber
& & \sum_{\chi } \frac{(-1)^{\chi}}
{(J-M-\chi )! (J+K-\chi )! (\chi )! (\chi+M-K)! } \\  \nonumber
& & [cos(\theta/2)]^{2J-2\chi+K-M)} [-sin(\theta/2)]^{2\chi+M-K}
\end{eqnarray}
	where the sum over $\chi$ is restricted to those values that do
not lead to negative factorials.

\appendix

{\bf Appendix C. Obtaining the Euler angles of a water molecule
from the Cartesian coordinates of its interaction sites} \\

	In our implementation of the MC algorithm we used Cartesian coordinates.
This is convenient when it comes to computing the
potential energy between two water molecules.
However, to evaluate the density function of the quantum 
rotational energy, the orientation
of the molecules must be specified in terms of  Euler angles. Therefore, we
need a procedure
to obtain the Euler angles that define the orientation of a given
molecule from the Cartesian coordinates of its
interaction sites. We used Euler angles $(\theta, \phi, \chi )$
as defined in Ref.\onlinecite{levine}; 
$\theta $ varies from 0 to $\pi$ and
$\phi $ and $\chi $ go from 0 to $2\pi $.
Let us denote $X,Y,Z$ as the orthogonal axes of a laboratory frame fixed in the space.
Let us assume that we associated three orthogonal axes to the molecule, namely $x,y,z$, which 
define the body frame (with its origin located at the centre of mass of the molecule). 
We shall assume that both sets of orthogonal axes are right handed. 
The orientation of the molecule can be defined by 
three Euler angles. The three Euler angles are defined by the operations required 
to move the molecule from an initial orientation, where $x,y,z$ are coincident with   
$X,Y,X$ (having both set of axes a common origin) to its current configuration (where  
the two set of axes also have a common origin). 
Rotations are counterclockwise. 
First, a rotation is performed about the $Z$-axis by an angle $\phi$,
so that the axes $X,Y,Z$ change to $x', y', Z$.
Secondly, we perform a rotation by an angle $\theta $ about the $y'$-axis obtained
from the previous rotation ($x',y',Z  \to x'',y',z'$).
Finally, a new rotation by an angle $\chi$ is performed around the $z'$-axis 
of the frame fixed in the body ( $x'',y',z'$ to $x,y,z$).
Using this convention, the coordinates in laboratory frame of a site of the 
molecule, $X,Y,Z$ (i.e., {\bf R} ) can be obtained easily from the coordinates of that site in 
body frame, x$_{b}$,y$_{b}$,z$_{b}$ (i.e., ${\bf r_{b}}$ )  using the expression :

\begin{equation}
\begin{pmatrix}
X \\
Y \\
Z
\end{pmatrix}
=
\begin{pmatrix}
\cos \phi \cos \theta \cos \chi - \sin \phi \sin \chi  & &
-\cos \phi \cos \theta \sin \chi - \sin \phi \cos \chi & &
\cos \phi \sin \theta \\
\sin \phi \cos \theta \cos \chi + \cos \phi \sin \chi & &
-\sin \phi \cos \theta \sin \chi + \cos \phi \cos \chi & &
\sin \phi \sin \theta \\
-\sin \theta \cos \chi & & 
\sin \theta \sin \chi & &
\cos \theta 
\end{pmatrix}
\begin{pmatrix}
x_{b}  \\
y_{b}  \\
z_{b}
\end{pmatrix}
\label{rotation_matrix}
\end{equation}
        Let us assign a particular body frame to the molecule of water. 
Taking the $bca$ convention, the body frame of the water molecule is chosen
so that the $b$ principle axis (i.e., that associated with $I_b$) is assigned 
to $x$, $c$ is assigned $y$ and  $a$ is assigned to $z$ (remember that
$I_a \leq I_b \leq I_c$).
Under this convention, the $b$ axis lies along the H-O-H bisector, the $c$ axis
is perpendicular to the molecular plane and the $a$ axis is parallel to the line
connecting the two hydrogen atoms. With this choice 
the coordinates of the oxygen are $(\alpha, 0,0)$ (the centre of mass in water 
is located upon the H-O-H bisector, slightly below the oxygen atom, and we shall define
the $x$ direction such that $\alpha$ is positive). The coordinates of the 
hydrogen atoms are be given by 
$(\beta , 0, -\gamma )$ and $(\beta , 0, \gamma )$. Notice that the value of $\beta $ 
should be negative, since $\alpha$ was taken to be positive. 
We denote $(X_O, Y_O, Z_O)$,
$(X_{H_1}, Y_{H_1}, Z_{H_1})$ and $(X_{H_2}, Y_{H_2}, Z_{H_2})$ 
as the coordinates of the oxygen and of the hydrogens, respectively, 
in the laboratory frame, whose origin is fixed to be the centre of mass of the molecule. As mentioned before, the centre
of mass of the molecule coincides with the origin of the body frame fixed
in the molecule.
For the oxygen, the coordinates in the body frame fixed in the molecule 
$(\alpha ,0,0)$ are related to those of the laboratory frame 
through 
Eq. \ref{rotation_matrix}. The multiplication
of matrices in Eq. \ref{rotation_matrix} leads to three equations:
\begin{equation}
X_O=\alpha ( \cos \phi \cos \theta \cos \chi - \sin \phi \sin \chi)
\label{xo}
\end{equation}
\begin{equation}
Y_O=\alpha ( \sin \phi \cos \theta \cos \chi + \cos \phi \sin \chi)
\label{yo}
\end{equation}
\begin{equation}
Z_O=- \alpha \sin \theta \cos \chi
\label{zo}
\end{equation}

	Analogously, the coordinates of the hydrogens in the laboratory frame 
are given by:
\begin{equation}
X_{H_1} = \beta \left( \cos \phi \cos \theta \cos \chi - \sin \phi \sin \chi \right)
- \gamma \cos \phi \sin \theta 
\label{xh1}
\end{equation}
\begin{equation}
Y_{H_1} = \beta \left( \sin \phi \cos \theta \cos \chi + \cos \phi \sin \chi \right)
- \gamma \sin \phi \sin \theta 
\label{yh1}
\end{equation}
\begin{equation}
Z_{H_1} = -\beta \sin \theta \cos \chi 
- \gamma \cos \theta 
\label{zh1}
\end{equation}
\begin{equation}
X_{H_2} = \beta \left( \cos \phi \cos \theta \cos \chi - \sin \phi \sin \chi \right)
+ \gamma \cos \phi \sin \theta 
\label{xh2}
\end{equation}
\begin{equation}
Y_{H_2} = \beta \left( \sin \phi \cos \theta \cos \chi + \cos \phi \sin \chi \right)
+ \gamma \sin \phi \sin \theta 
\label{yh2}
\end{equation}
\begin{equation}
Z_{H_2} = -\beta \sin \theta \cos \chi 
+ \gamma \cos \theta 
\label{zh2}
\end{equation}
	In summary, we have obtained nine equations to determine the three Euler angles
($\phi $ and $\chi $ go from 0 to $2\pi $ and, therefore, their 
value can only be unambiguously obtained from the knowledge of
both their sine and cosine; whereas $\theta $ varies from 0 to
$\pi $ so that it is only necessary to know its cosine).
The Euler angles for an instantaneous configuration where the
atomic coordinates are $(X_0, Y_0, Z_0)$, 
$(X_{H_1}, Y_{H_1}, Z_{H_1})$ and $(X_{H_2}, Y_{H_2}, Z_{H_2})$
can therefore be obtained by solving the set of equations defined
by Eqs. \ref{xo} - \ref{zh2}.
Subtracting Eq. \ref{zh2} from \ref{zh1} we obtain:
\begin{equation}
\cos \theta = \frac{ Z_{H2} - Z_{H1}}{2\gamma }
\end{equation}
	Subtracting Eqs.\ref{xh2} from \ref{xh1}:
\begin{equation}
\cos \phi = \frac{ (X_{H_2}-X_{H_1}) }{ 2 \gamma \sin \theta }
\end{equation}
	and subtracting Eqs. \ref{yh2} from \ref{yh1}:
\begin{equation}
\sin \phi = \frac{ (Y_{H_2}-Y_{H_1}) }{ 2 \gamma \sin \theta }
\end{equation}
	Finally, the Euler angle $\chi $ can be obtained 
from Eq.\ref{zo}: 
\begin{equation}
\cos \chi = \frac{-Z_{O}}{\alpha \sin \theta } 
\end{equation}
	and adding Eq.\ref{xh1} and Eq.\ref{xh2}:
\begin{equation}
\sin \chi = \left( \cos \phi \cos \theta \cos \chi -
	\frac { (X_{H_1}+X_{H_2}) }{ 2 \beta } \right)
	\frac{1}{\sin \phi}
\end{equation}
	or, alternatively, adding Eq.\ref{yh1} and Eq. \ref{yh2}:
\begin{equation}
\sin \chi = \left( -\sin \phi \cos \theta \cos \chi +
	\frac { (Y_{H_1}+Y_{H_2}) }{ 2 \beta } \right)
	\frac{1}{\cos \phi}
\end{equation}
	In the special case that $\theta =0$  or $\theta =\pi$
the expressions given above are not valid because the denominator
vanishes, resulting in a singularity. The probability of obtaining
exactly $\theta =0$  or $\theta =\pi$ is very small during a simulation.
In these special cases, alternative
expressions can be obtained by evaluating of the rotation matrix
for the particular value of $\theta $.
When $\theta =0$, $\sin \theta = 0$ and $\cos \theta = 1$ and,
therefore, the rotation matrix (Eq. \ref{rotation_matrix}) becomes
\begin{equation}
\begin{pmatrix}
\cos (\phi + \chi) & &
-\sin (\phi + \chi ) & & 
0 \\
\sin (\phi + \chi ) & &
\cos (\phi + \chi ) & & 
0 \\
0 & &
0 & &
1
\end{pmatrix}
\label{rotation_matrixb}
\end{equation}
i.e, in this case, the rotation can be seen as a simple rotation about the
$z$-axis by an angle $\phi'=\phi + \chi $. In the case $\theta=0$ there is 
no a unique way of defining $\phi$ and $\chi$ individually, as any combination 
of $\phi$ and  $\chi$ having the same value of $\phi'$ ($\phi'=\phi + \chi $) will 
provide the same final configuration. Here for the particular case $\theta=0$ we 
decided to assign $\phi=0$ , and with this choice 
the sine and cosine of the $\chi$ angle 
can be readily obtained using a procedure similar to that outlined
above for a general case but using the rotation matrix given 
in Eq. \ref{rotation_matrixb}. Using this procedure we obtain :
\begin{equation}
\sin (\chi)=\frac{Y_{H_1}}{\beta }
\end{equation}
and
\begin{equation}
\cos (\chi)=\frac{X_{H_1}}{\beta }
\end{equation}
	Finally, when $\theta =\pi$, $\sin \theta =0$ and $\cos \theta = -1 $
so that the rotation matrix is now:
\begin{equation}
\begin{pmatrix}
-\cos (\chi - \phi) & &
\sin (\chi - \phi ) & & 
0 \\
\sin (\chi - \phi ) & &
\cos (\chi - \phi ) & & 
0 \\
0 & &
0 & &
-1
\end{pmatrix}
\label{rotation_matrixb_2}
\end{equation}
	In this case, this rotation can be seen as a simple rotation
about the $z$-axis  by an angle $\phi'=\chi -\phi $. Again it is not possible 
to assign in a unique way values of $\chi$ and $\phi$. For this reason we 
arbitrarily assigned in this case $\phi=0$ so that $\chi$ is obtained as:
\begin{equation}
\sin \chi = \frac{ Y_{H_1}}{\beta }
\end{equation}
\begin{equation}
\cos \chi = - \frac{ X_{H_1}}{\beta }
\end{equation}

\appendix

{\bf Appendix D. Obtaining the relative Euler angles of replica $t+1$
with respect to those of replica $t$.} \\ 

  Let us focus on two replicas $t$ and $t+1$ of a certain molecule.
	Suppose that ${\bf R}^{t}$ are the laboratory frame 
instantaneous coordinates of a certain 
site of replica $t$ with respect to its centre of mass (the centre of mass of replica 
$t$). These coordinates can be obtained from the body frame coordinates of that site 
(${\bf r}_{b}$) 
using the rotation matrix $M_{t}$:
\begin{equation}
{\bf R}^{t} = M_{t} {\bf r}_{b}
\end{equation}
Inverting the previous equation one obtains
\begin{equation}
{\bf r}_{b} = {M_{t}}^{-1} {\bf R}^{t}
\end{equation}
	The relative Euler angles of replica $t+1$ with respect to
those of replica $t$ can be computed by expressing the 
instantaneous coordinates of replica $t+1$ (${\bf R}^{t+1}$) in the
reference system of replica $t$  by:
\begin{equation}
{\bf R'}^{t+1}= {M_{t}}^{-1} {\bf R}^{t+1}
\end{equation}
In other words, replica $t+1$ is rotated using the rotation matrix 
${M_{t}}^{-1}$ to obtain its orientation with respect to that of replica $t$.
The relative orientation of molecule $t+1$ with respect to that of molecule
$t$ is given by the atomic coordinates ${{\bf R'}}^{t+1}$. The Euler angles
associated with this orientation, i.e., the Euler angles of replica $t+1$ with
respect to those of replica $t$, can be computed using the procedure described in
Appendix C.

\end{document}